\documentclass[a4paper,fleqn,usenatbib]{mnras}

\usepackage{newtxtext,newtxmath}
\usepackage[T1]{fontenc}

\usepackage[english]{babel}
\usepackage{graphicx}
\usepackage{amsmath}
\usepackage{booktabs}
\usepackage{hyperref}
\usepackage{breakcites}

\usepackage{comment}
\usepackage{enumitem}
\usepackage{breqn}

\setlist[enumerate]{
  labelsep=8pt,
  labelindent=0.\parindent,
 itemindent=0pt,
  leftmargin=*,
}

\usepackage{color, colortbl}

\newcommand*{\mycdot}{\kern-.2em\cdot\kern-.2em}
\renewcommand{\S}{Section}
\newcommand{\Ss}{Sections}
\newcommand{\F}{Fig.}
\newcommand{\Fs}{Figs}
\newcommand{\Eq}{Equation}
\newcommand{\Eqs}{Equations}

\newcommand{\epsoct}{\epsilon_\mathrm{oct}}

\newcommand{\TZ}{\Theta}
\newcommand{\TZz}{\Theta_0}
\newcommand{\CZ}{\varepsilon}
\newcommand{\CZz}{\varepsilon_0}

\newcommand{\FP}{\mathrm{FP}}
\newcommand{\crit}{\mathrm{crit}}
\newcommand{\rel}{\mathrm{rel}}
\newcommand{\ZLK}{\mathrm{ZLK}}

\newcommand{\peak}{\mathrm{peak}}
\newcommand{\stab}{\mathrm{stab}}
\newcommand{\DA}{\mathrm{DA}}

\usepackage{listings}

\usepackage{color}

\definecolor{dkgreen}{rgb}{0,0.6,0}
\definecolor{gray}{rgb}{0.5,0.5,0.5}
\definecolor{mauve}{rgb}{0.58,0,0.82}

\allowdisplaybreaks

\begin{document}
\title[ZLK in the non-test-particle limit]{Properties of von Zeipel-Lidov-Kozai oscillations in triple systems at the quadrupole order: relaxing the test particle approximation}

\author[Hamers]{Adrian S. Hamers$^{1}$\thanks{E-mail: hamers@mpa-garching.mpg.de} \\
$^{1}$Max-Planck-Institut f\"{u}r Astrophysik, Karl-Schwarzschild-Str. 1, 85741 Garching, Germany}
\date{Accepted 2020 November 6. Received 2020 October 21; in original form 2020 August 3}

\label{firstpage}
\pagerange{\pageref{firstpage}--\pageref{lastpage}}
\maketitle

\begin{abstract}  
Von Zeipel-Lidov-Kozai (ZLK) oscillations in hierarchical triple systems have important astrophysical implications such as triggering strong interactions and producing, e.g., Type Ia supernovae and gravitational wave sources. When considering analytic properties of ZLK oscillations at the lowest (quadrupole) expansion order, as well as complications due to higher-order terms, one usually assumes the test particle limit, in which one of the bodies in the inner binary is massless. Although this approximation holds well for, e.g., planetary systems, it is less accurate for systems with more comparable masses such as stellar triples. Whereas non-test-particle effects are usually taken into account in numerical simulations, a more analytic approach focusing on the differences between the test particle and general case (at quadrupole order) has, to our knowledge, not been presented. Here, we derive several analytic properties of secular oscillations in triples at the quadruple expansion order. The latter applies even to relatively compact triples, as long as the inner bodies are similar in mass such that octupole-order effects are suppressed. We consider general conditions for the character of the oscillations (circular versus librating), minimum and maximum eccentricities, and timescales, all as a function of $\gamma \equiv (1/2) \, L_1/G_2$, a ratio of inner-to-outer orbital angular momenta variables ($\gamma=0$ in the test particle limit). In particular, eccentricity oscillations are more effective at retrograde orientations for non-zero $\gamma$; assuming zero initial inner eccentricity, the maximum eccentricity peaks at $\cos(i_{\rel,\,0}) = -\gamma$, where $i_{\rel,\,0}$ is the initial relative inclination. We provide a \textsc{Python} script which can be used to quickly compute these properties.
\end{abstract}


\section{Introduction}
\label{sect:introduction}
Hierarchical triple systems are well known for their rich dynamics which can have important astrophysical implications. If the inner and outer orbits are initially mutually highly inclined, then the gravitational torque of the outer orbit can induce high-amplitude eccentricity oscillations in the inner binary, known as Lidov-Kozai (LK) or von Zeipel-Lidov-Kozai (ZLK) oscillations (\citealt{1910AN....183..345V,1962P&SS....9..719L,1962AJ.....67..591K}; see \citealt{2016ARA&A..54..441N,2017ASSL..441.....S,2019MEEP....7....1I} for reviews). The high eccentricities attained during these oscillations can give rise to strong interactions. For example, they can help to produce short-period binaries (e.g., \citealt{1979A&A....77..145M,1998MNRAS.300..292K,2001ApJ...562.1012E,2006Ap&SS.304...75E,2007ApJ...669.1298F,2014ApJ...793..137N,2018MNRAS.479.4749B,2019MNRAS.488.2480R}) and hot Jupiters (e.g., \citealt{2003ApJ...589..605W,2007ApJ...669.1298F,2012ApJ...754L..36N,2015ApJ...799...27P,2016MNRAS.456.3671A,2016ApJ...829..132P}), enhance mergers of compact objects (e.g., \citealt{2002ApJ...578..775B,2011ApJ...741...82T,2013MNRAS.430.2262H,2017ApJ...841...77A,2017ApJ...836...39S,2017ApJ...846L..11L,2018ApJ...863...68L,2018ApJ...865....2H,2018ApJ...856..140H,2018ApJ...853...93R,2018ApJ...864..134R,2018A&A...610A..22T,2019MNRAS.486.4443F}), affect the evolution of protoplanetary or accretion disks in binaries (e.g., \citealt{2014ApJ...792L..33M,2015ApJ...813..105F,2017MNRAS.467.1957Z,2017MNRAS.469.4292L,2018MNRAS.477.5207Z,2019MNRAS.485..315F,2019MNRAS.489.1797M}), trigger white dwarf pollution by planets (e.g., \citealt{2016MNRAS.462L..84H,2017ApJ...834..116P}), and produce blue straggler stars (e.g., \citealt{2009ApJ...697.1048P,2016ApJ...816...65A,2016MNRAS.460.3494S,2019MNRAS.488..728F}). Also, in stellar triples, ZLK oscillations can combine with stellar evolution to trigger interactions during or after the main sequence (e.g., \citealt{2013MNRAS.430.2262H,2013ApJ...766...64S,2014ApJ...794..122M,2016ComAC...3....6T,2016MNRAS.460.3494S,2017ApJ...841...77A,2018A&A...610A..22T,2019ApJ...878...58S,2019MNRAS.489.2298C,2019ApJ...882...24H}). 

The theory of ZLK oscillations has a rich history. Usually, an expansion of the Hamiltonian is made in the ratio of separations of the inner and outer orbits. The expanded Hamiltonian can be expressed in canonical Delaunay orbital elements. After averaging the Hamiltonian over both orbits (also known as the von Zeipel transformation, \citealt{1959AJ.....64..378B}), the dependence on the orbital phases is removed, such that the associated conjugate momenta of the inner and outer orbits, $L_1$ and $L_2$, are conserved. Since $L_1 \propto \sqrt{a_1}$ and $L_2 \propto \sqrt{a_2}$, where $a_1$ and $a_2$ are the inner and outer semimajor axes, respectively, this implies that the semimajor axes are conserved as well. Hamilton's equations then give the equations of motion. 

Lidov \citep{1962P&SS....9..719L} and Kozai \citep{1962AJ.....67..591K} studied the problem of satellites around the Moon and Jupiter, respectively. For their applications, it was well justified to assume the test particle limit, in which the angular momentum of the inner orbit is negligible. The latter is the case, for example, if one of the bodies in the inner orbit has a negligible mass (or is completely massless)\footnote{A variation of the test particle limit discussed here is when the outer object is massless; see \citet{2017AJ....154...18N,2018MNRAS.474.4855V,2019A&A...627A..17D}.}. The test particle limit is very useful since it simplifies the equations of motion and properties of ZLK oscillations. In particular, in the test particle approximation, $\sqrt{1-e_1^2} \cos(i_\rel)$ is constant, where $e_1$ is the inner orbit eccentricity, and $i_\rel$ is the relative inclination between the inner and outer orbits. This implies a simple relation between eccentricity and inclination, and, in the case of zero initial inner orbit eccentricity, leads to the simple canonical relation $e_\max = \sqrt{1-(5/3) \cos^2 (i_\rel)}$ for the maximum inner eccentricity reached during ZLK oscillations.

When considering analytic properties of ZLK oscillations, the test particle assumption is usually made (e.g., \citealt{1999CeMDA..75..125K,2007CeMDA..98...67K,2007ApJ...669.1298F,2015MNRAS.452.3610A,2020MNRAS.tmp.2090H}). However, as pointed out by \citet{2013MNRAS.431.2155N}, $\sqrt{1-e_1^2} \cos(i_\rel)$ is no longer conserved in the non-test-particle case and this can give rise to flips in the orbital orientation even at the lowest (quadrupole) expansion order, and not requiring the inclusion of octupole-order terms (e.g., \citealt{2011ApJ...742...94L,2011PhRvL.107r1101K,2013ApJ...779..166T,2014ApJ...791...86L}), or the presence of additional bodies (e.g., \citealt{2013MNRAS.435..943P,2015MNRAS.449.4221H,2017MNRAS.470.1657H,2018MNRAS.474.3547G}). This problem conspires with the fact that, when expressed in canonical Delaunay variables in which the reference frame is the invariable plane (perpendicular to the total angular momentum), the Hamiltonian depends on the longitudes of the ascending nodes, $h_j$, only through the combination $\Delta h = h_1-h_2=\pi$. When substituting this relation, the Hamiltonian becomes independent of $h_j$, giving the (incorrect) impression that the conjugate momentum of $h_j$ should be constant, and hence as well the $z$-components of the angular momenta \citep{2013MNRAS.431.2155N}. 

In subsequent studies, these complications have usually been taken into account when numerically integrating the equations of motion. Specifically, one should not compute the relative inclination from an assumed-to-be constant $\sqrt{1-e_1^2} \cos(i_\rel)$, but from conservation of the total angular momentum. Alternatively, the Hamiltonian and the equations of motion can be formulated entirely in vector form (e.g., \citealt{2015ApJ...799...27P,2015MNRAS.447..747L,2016MNRAS.459.2827H}), in which case this problem is circumvented altogether. 

Nevertheless, it is useful to consider, using more analytic methods, properties of ZLK oscillations in the non-test-particle limit. The non-test-particle limit (also known as the `non-restricted' or `stellar' limit), was considered with analytic methods by \citet{1968AJ.....73..190H}, \citet{1969CeMec...1..200H}, \citet{1976CeMec..13..471L}, and \citet{1994CeMDA..58..245F}. However, these authors considered the problem quite abstractly and qualitatively, with little emphasis on the implications of taking the general limit (as opposed to test particle limit), and bearing in mind astrophysically relevant quantities such as the maximum eccentricities, and eccentricity oscillation timescales. Also, we note that \citet{2013MNRAS.431.2155N} derived an expression for the maximum eccentricity in the non-test-particle limit (see Appendix A and Equation A42 therein); the general equations were included, but further analysis (beyond the assumption of zero initial inner and outer eccentricities) was not carried out. Some aspects of the non-test-particle limit were also considered by \citet{2017MNRAS.467.3066A}.

Here, we present a detailed study of the maximum eccentricities and eccentricity oscillation timescales using both analytic and semianalytic methods. Our results are useful for interpreting data from numerical studies. For example, in population synthesis studies of stellar triples, the initial relative inclination distributions for merging/strongly interacting systems are often found to be peaked around a value slightly above $90^\circ$ (e.g., \citealt{2013MNRAS.430.2262H}). This cannot be explained by canonical results that apply to the test particle limit. In particular, we will show that, in the general case (but still assuming the quadrupole expansion order), the maximum eccentricity instead peaks at $\cos(i_{\rel,\,0}) = -\gamma$, where $\gamma \equiv (1/2) \, L_1/G_2$ is a ratio of inner-to-outer orbital angular momenta variables ($\gamma=0$ in the test particle limit). Furthermore, we will show that the maximum eccentricity can either increase or decrease as a function of $\gamma$ depending on the initial relative orientation, and that these trends are also reflected in the timescales of the eccentricity oscillations. 

The structure of this paper is as follows. In \S~\ref{sect:tp}, we briefly give some basic background information on ZLK oscillations in the test particle limit, focusing on analytic properties (the reader well familiar with the topic may want to skip this section). We generalise this to the non-test-particle case in \S~\ref{sect:gen}. In \S~\ref{sect:gamma}, we explore the dependence of several important quantities (maximum eccentricities and timescales) as a function of $\gamma$. We discuss in \S~\ref{sect:discussion}, and conclude in \S~\ref{sect:conclusions}.

\section{Test particle limit}
\label{sect:tp}
\subsection{Preliminaries}
\label{sect:tp:pre}
Consider a hierarchical triple with inner masses $m_1$ and $m_2$ ($m_2 \leq m_1$), and an outer third body with mass $m_3$. The inner and outer orbits have semimajor axes and eccentricities denoted with $a_1$ and $a_2$ and $e_1$ and $e_2$, respectively. For notational convenience, we let $e\equiv e_1$, and $x\equiv 1 - e^2$. Let the relative inclination between the orbits be $i_{\mathrm{rel}}$, and we define $\theta \equiv \cos(i_{\mathrm{rel}})$. The arguments of periapsis of the inner and outer orbit are $g_1\equiv g$ and $g_2$, respectively. 

The double-averaged Hamiltonian, to quadrupole expansion order, is given by (e.g., \citealt{2000ApJ...535..385F,2002ApJ...578..775B,2013MNRAS.431.2155N})
\begin{align}
\label{eq:H}
H_2 = C_2 \left [ \left (2+3e^2 \right ) \left (3\theta^2 - 1 \right ) + 15 e^2 \left(1-\theta^2 \right ) \cos(2g) \right ],
\end{align}
where
\begin{align}
\label{eq:C2}
C_2 \equiv \frac{1}{16} \frac{\mathcal{G} m_1m_2m_3}{(m_1+m_2)a_2} \left ( \frac{a_1}{a_2} \right )^2 \left (1-e_2^2 \right )^{-3/2},
\end{align}
and with $\mathcal{G}$ the gravitational constant. Since $H_2$ does not depend on $g_2$, $G_2$ (see below) is constant, and hence $e_2$ is constant as well (a `happy coincidence' according to \citealt{1976CeMec..13..471L}). This is the case irrespective of whether or not the test particle approximation is made. Furthermore, since the orbital phases have been eliminated by double averaging, $a_1$ and $a_2$ are constant as well, such that $C_2$ is constant. Conservation of energy (within the approximations of double averaging and truncating the expansion at the quadrupole order) implies that $H_2$ is conserved.

Another conserved quantity is the total orbital angular momentum (we do not consider spin angular momenta). Let $L_j = \mu_j \sqrt{\mathcal{G} M_j a_j}$ be the circular angular momenta (conjugate momentum to the orbital phase), with $M_1 \equiv m_1+m_2$ and $M_2 \equiv M_1+m_3$; the reduced masses are $\mu_1 = m_1 m_2/M_1$, and $\mu_2 = M_1 m_3/M_2$. Also, the non-circular angular momenta (conjugate to $g_j$) are $G_j = L_j \sqrt{1-e_j^2}$. The squared total angular momentum is then given by
\begin{align}
G_{\mathrm{tot}}^2 = L_1^2 \left(1-e^2 \right) +L_2^2 \left(1-e_2^2 \right) + 2 L_1 L_2 \sqrt{1-e^2} \sqrt{1-e_2^2} \theta.
\end{align}
Indicating initial variables with a subscript `0', conservation of orbital angular momentum (combined with the fact that $e_2$ is constant within out approximation) implies
\begin{align}
\label{eq:theta}
\theta = \frac{\TZ - \gamma \left(1-e^2 \right )}{\sqrt{1-e^2}},
\end{align}
where
\begin{align}
\label{eq:tz}
\TZ \equiv \sqrt{1-e_0^2} \, \theta_0 + \gamma \left (1-e_0^2 \right ),
\end{align}
and with the constant parameter
\begin{align}
\label{eq:gammadef}
\nonumber \gamma &\equiv \frac{1}{2} \frac{1}{\sqrt{1-e_2^2}} \frac{L_1}{L_2} = \frac{1}{2} \frac{L_1}{G_2} \\
\quad &= \frac{1}{2}  \frac{1}{\sqrt{1-e_2^2}} \left ( \frac{a_1}{a_2} \right )^{1/2} \frac{m_1 m_2}{(m_1+m_2)m_3} \left ( \frac{m_1+m_2+m_3}{m_1+m_2} \right )^{1/2}.
\end{align}
If one of the inner binary bodies is massless, then $\gamma=0$. \Eq~(\ref{eq:theta}) then reduces to
\begin{align}
\label{eq:thetatp}
\sqrt{1-e^2} \, \theta = \sqrt{1-e_0^2} \, \theta_0 \equiv \TZz = \mathrm{constant}.
\end{align}
This is the well known `test particle' limit, in which the $z$-component of the inner binary angular momentum (also referred to as the `Kozai' or `ZLK constant') is conserved. 

The equations of motion follow from Hamilton's equations. These read, for $e$ and $g$,
\begin{subequations}
\begin{align}
\label{eq:eome}
\frac{\mathrm{d} e}{\mathrm{d} t} &= C_2 \frac{\sqrt{1-e^2}}{L_1} 30 e \left (1-\theta^2 \right ) \sin(2g); \\
\label{eq:eomg}
\nonumber \frac{\mathrm{d} g}{\mathrm{d} t} &= 6 \, C_2 \Biggl \{ \frac{1}{L_1 \sqrt{1-e^2}} \left [ 4 \theta^2 + (5 \cos(2g) - 1) \left (1-e^2 - \theta^2 \right ) \right ] \\
&\quad + \frac{\theta}{G_2} \left [2 + e^2 \left (3-5 \cos(2g) \right ) \right ] \Biggl \}.
\end{align}
\end{subequations}
The inclination $\theta$ at any time follows from \Eq~(\ref{eq:theta}). We do not give the equations for motion for $h_1$ and $h_2$ since they are decoupled from the Hamiltonian (see, e.g., \citealt{2013MNRAS.431.2155N}).

\begin{figure}
\center
\includegraphics[scale = 0.45, trim = 8mm 0mm 8mm 0mm]{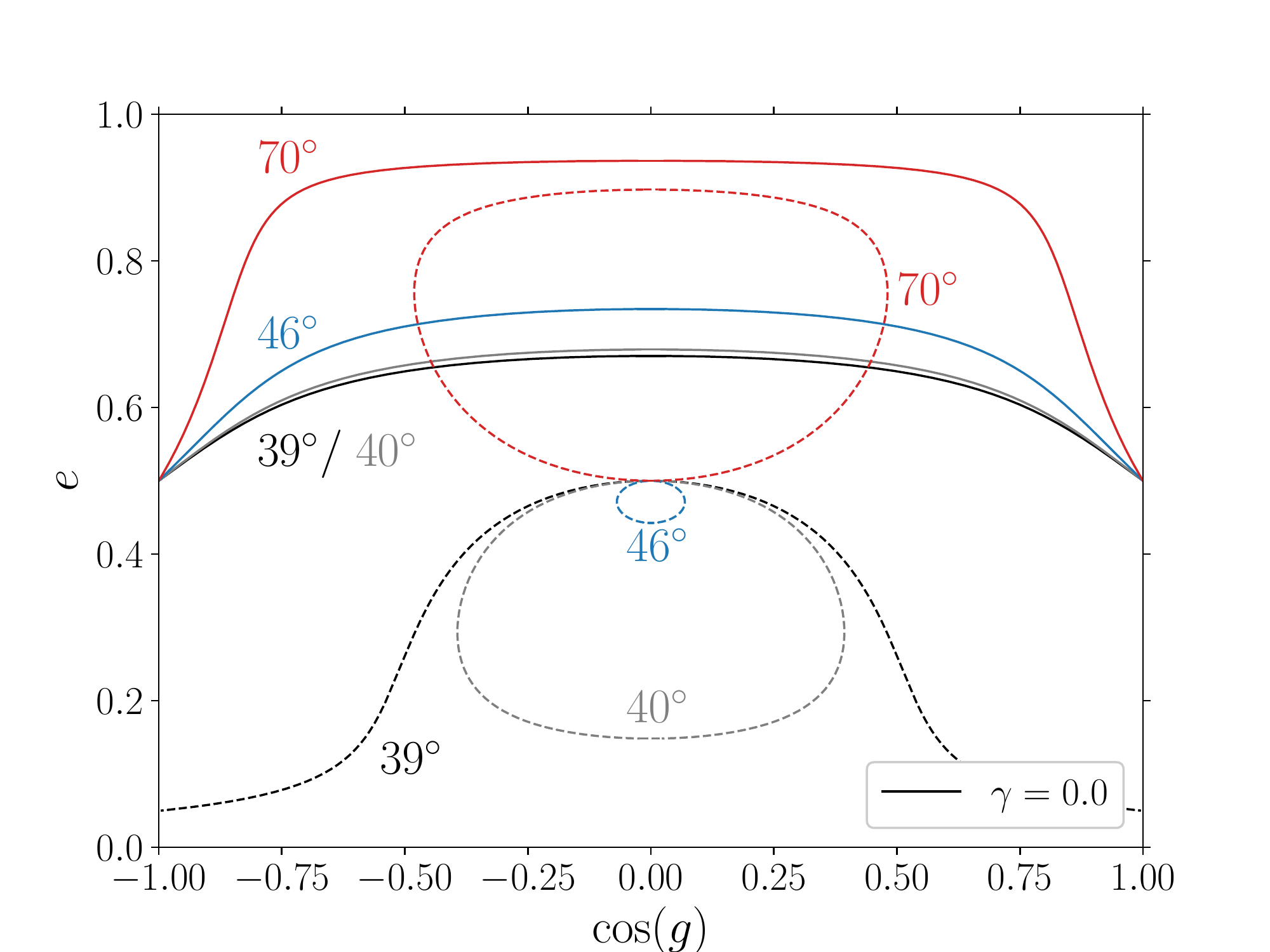}
\caption{Phase space plot of $(\cos g,e)$, where $g$ and $e$ are the inner orbit argument of periapsis and eccentricity, respectively, for a triple in the test particle limit with initial values $e_0=0.5$, $g_0=0$ (solid lines) or $g_0=\pi/2$ (dashed lines), and several values of the initial inclination (indicated with different colours). We include solutions with $i_{\rel,\,0} = 39^\circ$ (black), $i_{\rel,\,0} = 40^\circ$ (grey), $i_{\rel,\,0} = 46^\circ$ (blue), and $i_{\rel,\,0} = 70^\circ$ (red). }
\label{fig:cont}
\end{figure}

\subsection{Phase space plots}
\label{sect:tp:cont}
It is instructive to consider phase space plots in the $(\cos g,e)$ parameter space. From \Eq~(\ref{eq:H}), the inner argument of periapsis is given by
\begin{align}
\label{eq:cos2g}
\cos 2g = \frac{c_0 - \left (2+3e^2 \right ) \left (3 \theta^2 -1 \right )}{15 e^2 \left (1-\theta^2 \right )},
\end{align}
with $\theta$ given by \Eq~(\ref{eq:theta}). Here, $c_0$ is the initial Hamiltonian normalised by $C_2$, i.e.,
\begin{align}
c_0 \equiv \left (2+3e_0^2 \right ) \left (3\theta_0^2 - 1 \right ) + 15 e_0^2 \left(1-\theta_0^2 \right ) \cos(2g_0).
\end{align}
\F~\ref{fig:cont} shows contours in the $(\cos g,e)$ plane assuming $\gamma=0$ (test particle limit) and $e_0=0.5$, for several initial values of $\theta_0$ (corresponding to different colors), and $g_0=0$ (solid lines) and $g_0=\pi/2$ (dashed lines). 

For $g_0=0$, oscillations `circulate' between $\cos g = 1$ and $\cos g = -1$. The minimum and maximum eccentricities are reached at $g=k \pi$ and $g = (1/2 +k) \pi$, respectively, where $k\in \mathbb{Z}$. 

For $g_0=\pi/2$, $g$ remains bound between two values symmetrically around $g=\pi/2$, and `librates' between them. The eccentricity minima and maxima are both reached at $g = \pi/2$. In this case of librating oscillations, the maximum eccentricity exceeds the initial eccentricity if the inclination is sufficiently large (see the curve corresponding to $70^\circ$). As the initial inclination is decreased, the maximum eccentricity decreases whereas the minimum remains fixed at $e_0=0.5$, i.e., the size of the `libration island' decreases. At a certain critical inclination, the libration curve reduces to a single point, known as the `fixed point'. For even smaller initial inclinations, the curves `flip' over, and the maximum eccentricity becomes $e_0$, whereas the minimum eccentricity is $<e_0$. Furthermore, once $i_{\mathrm{rel,\,0}}<40^\circ$, the libration curves `spread out' and switch to circulating oscillations. 

Generally, whether the oscillation will be circulating or librating given the initial conditions can be understood from \Eq~(\ref{eq:cos2g}). The boundary between circulating and librating solutions is given by the curve which reaches $e=0$ at $\cos g = \pm1$ (compare the two dashed curves in \F~\ref{fig:cont} with $i_{\rel,\,0} = 39^\circ$ and $i_{\rel,\,0} = 40^\circ$). This condition is equivalent to $\cos2g = 1$ with $e=0$; setting $\gamma=0$ and applying \Eq~(\ref{eq:cos2g}), it translates into the condition $\CZz = 0$, where we define
\begin{align}
\label{eq:czz}
\CZz \equiv \frac{1}{12} \left ( c_0 - 6 \TZz^2 + 2 \right ).
\end{align}
If $\CZz>0$, the solution circulates, and it librates if $\CZz<0$. Note that the factor of $1/12$ in \Eq~(\ref{eq:czz}) is included to be consistent with the corresponding quantity originally defined by \citet{1962P&SS....9..719L}. In particular, substituting the definitions of $c_0$ and $\TZz$, \Eq~(\ref{eq:czz}) can be written as
\begin{align}
\label{eq:czz2}
\CZz = e_0^2 \left [1-\frac{5}{2} \left(1-\theta_0^2\right ) \sin^2(g_0) \right ].
\end{align}
This also shows that, when $e_0=0$, this always corresponds exactly to the boundary between circulating and librating orbits. 

Furthermore, we can obtain the location of the fixed point by setting $\dot{g}=0$ in \Eq~(\ref{eq:eomg}) with $g=\pi/2$, and $e=e_0$. In the test particle limit, $G_2 \gg L_1$ so we can neglect the term $\propto 1/G_2$ in  \Eq~(\ref{eq:eomg}). This gives, for $\gamma=0$, 
\begin{align}
\label{eq:fptp}
\theta_{0,\,\FP} = \sqrt{ \frac{3}{5} \left (1-e_0^2 \right )}.
\end{align}
In \F~\ref{fig:cont}, this gives a critical inclination of the fixed point of $i_{\mathrm{rel,\,0,\,\FP}} \simeq 47.9^\circ$.

\subsection{Eccentricity extrema}
\label{sect:tp:ecc}
Extrema of the eccentricity oscillations can be found by choosing $g$ in \Eq~(\ref{eq:H}) corresponding to an eccentricity minimum or maximum, using \Eq~(\ref{eq:theta}) to express $\theta=\theta(e)$, and solving for $e$. In the test particle limit and assuming $e_0=0$, this gives the following canonical relation for the maximum eccentricity,
\begin{align}
\label{eq:xmincan}
x_\min = \frac{5}{3} \theta_0^2
\end{align}
(recall that we defined $x\equiv 1-e^2$). Note that there is no dependence in \Eq~(\ref{eq:xmincan}) on $g_0$ since the initial argument of periapsis is ill-defined/not applicable when $e_0=0$. Since $x_\min$ cannot be $\geq1$, this implies that eccentricity excitation (when $e_0=0$) can only occur if $\theta < \theta_\crit = \sqrt{3/5}$, corresponding to $i_{\mathrm{rel,\,\crit}} \simeq 39.23^\circ$ (or $140.77^\circ$ for retrograde orbits). Also, we remark that \Eq~(\ref{eq:xmincan}) is fully symmetric with respect to $\theta_0$: prograde orbits behave exactly the same as their equivalent retrograde orbits (i.e., $x_\min$ is invariant under $\theta_0 \rightarrow -\theta_0$). 

When $e_0\neq 0$, the expression for $x_\min$ is more complicated, as it now also depends on $e_0$ and $g_0$. Expressions for $x_\min$ and $x_\max$ were given by, e.g., \citet{2015MNRAS.452.3610A}; for the maximum eccentricity,
\begin{align}
\label{eq:xmintp}
x_\min = \frac{1}{6} \left ( \zeta - \sqrt{\zeta^2 - 60 \, \TZz^2} \right ).
\end{align}
The minimum eccentricity is given by
\begin{align}
\label{eq:xmaxtplib}
x_\max = \frac{1}{6} \left ( \zeta + \sqrt{\zeta^2 - 60 \, \TZz^2} \right ), \quad (\CZz<0)
\end{align}
for librating solutions, whereas
\begin{align}
\label{eq:xmaxtpcirc}
x_\max = 1-\CZz, \quad (\CZz>0)
\end{align}
for circulating solutions. Here,
\begin{align}
\zeta \equiv 3 + 5 \TZz^2 + 2 \CZz.
\end{align}
We remark that \citet{2015MNRAS.452.3610A} used slightly different notation; in particular, $\CZz$ in our notation corresponds to `$C_{\mathrm{KL}}$' of \citet{2015MNRAS.452.3610A}, whereas $\TZz$ in our notation corresponds to `$\sqrt{\Theta}$' of \citet{2015MNRAS.452.3610A}. Also, note that `$\Theta$' in \citet{2015MNRAS.452.3610A} was always positive, whereas the $\TZ$ defined in \Eq~(\ref{eq:tz}) can be negative if the orientation is initially retrograde ($\theta_0<0$).

\subsection{Timescales}
\label{sect:tp:time}
The timescale of the eccentricity oscillations (the time between, e.g., two eccentricity maxima) can be derived by integrating the equation of motion for $e$ (\Eq~\ref{eq:eome}), using \Eq~(\ref{eq:cos2g}) to express $g$ in terms of $e$ and $\theta$, and \Eq~(\ref{eq:theta}) to express $\theta$ in terms of $e$. The resulting equation of motion in terms of $x\equiv 1-e^2$ is
\begin{align}
\label{eq:xdot}
\nonumber \dot{x} &= \pm 4 \frac{C_2}{L_1} \sqrt{x} \sqrt{20 + c_0 - 30\theta^2 + 6x \left (4\theta^2-3 \right )} \\
&\qquad \times \sqrt{10-c_0+6x \left(\theta^2 -2 \right )},
\end{align}
where the $+$ and $-$ signs applies to the phases of decreasing and increasing eccentricity, respectively. By symmetry of ZLK oscillations (which also applies in the non-test-particle limit), one only needs to integrate over $x$ from $x_\min$ to $x_\max$ and multiply by a factor of two. Therefore,
\begin{align}
\label{eq:tzlk}
\nonumber T_\ZLK &= 2\int_{x_\min}^{x_\max} \, \frac{\mathrm{d} x}{|\dot{x}|} \\
\nonumber &= \frac{L_1}{C_2} \int_{x_\min}^{x_\max} \, \mathrm{d} x \frac{1}{2} \left \{ x \left [20 + c_0 - 30\theta^2 + 6x \left (4\theta^2-3 \right ) \right ] \right. \\
&\left. \quad \times \left [10-c_0+6x \left(\theta^2 -2 \right ) \right ] \right \}^{-1/2},
\end{align}
where, again, $\theta=\theta(x)$ is given by \Eq~(\ref{eq:theta}). The minimum and maximum values of $x$, $x_\min$ and $x_\max$, can be calculated analytically (cf. \S~\ref{sect:tp:ecc}). Note that $L_1/C_2$ can be written as
\begin{align}
\label{eq:l1divc2}
\frac{L_1}{C_2} = \frac{16}{2\pi} \frac{P_2^2}{P_1} \frac{m_1+m_2+m_3}{m_3} \left (1-e_2^2 \right )^{3/2},
\end{align}
where $P_1=2\pi \sqrt{a_1^3/(GM_1)}$ and $P_2=2\pi \sqrt{a_2^3/(GM_2)}$ denote the inner and outer Keplerian orbital periods, respectively. 

The integral in \Eq~(\ref{eq:tzlk}) is generally hard to solve in closed form. In the case $\gamma=0$, it can be written in terms of elliptic integrals. \citet{2015MNRAS.452.3610A} found a useful simple fitting formula to the integral in the case $\gamma=0$ (cf. their equation 48). We briefly mention that, in the test particle limit, the ZLK timescale does not vary by more than factors of a few unless close to the circulation/libration boundary (e.g., \citealt{2015MNRAS.452.3610A}).

\section{General case}
\label{sect:gen}
Having considered the test particle limit in \S~\ref{sect:tp}, we here generalise to the case when the inner orbit angular momentum is not negligible, i.e., when $\gamma \neq 0$. Energy conservation still implies that \Eq~(\ref{eq:H}) is constant, and angular-momentum conservation implies the relation \Eq~(\ref{eq:theta}) between initial parameters and the current eccentricity, where now $\gamma \neq 0$.  

\begin{figure}
\center
\includegraphics[scale = 0.45, trim = 8mm 0mm 8mm 0mm]{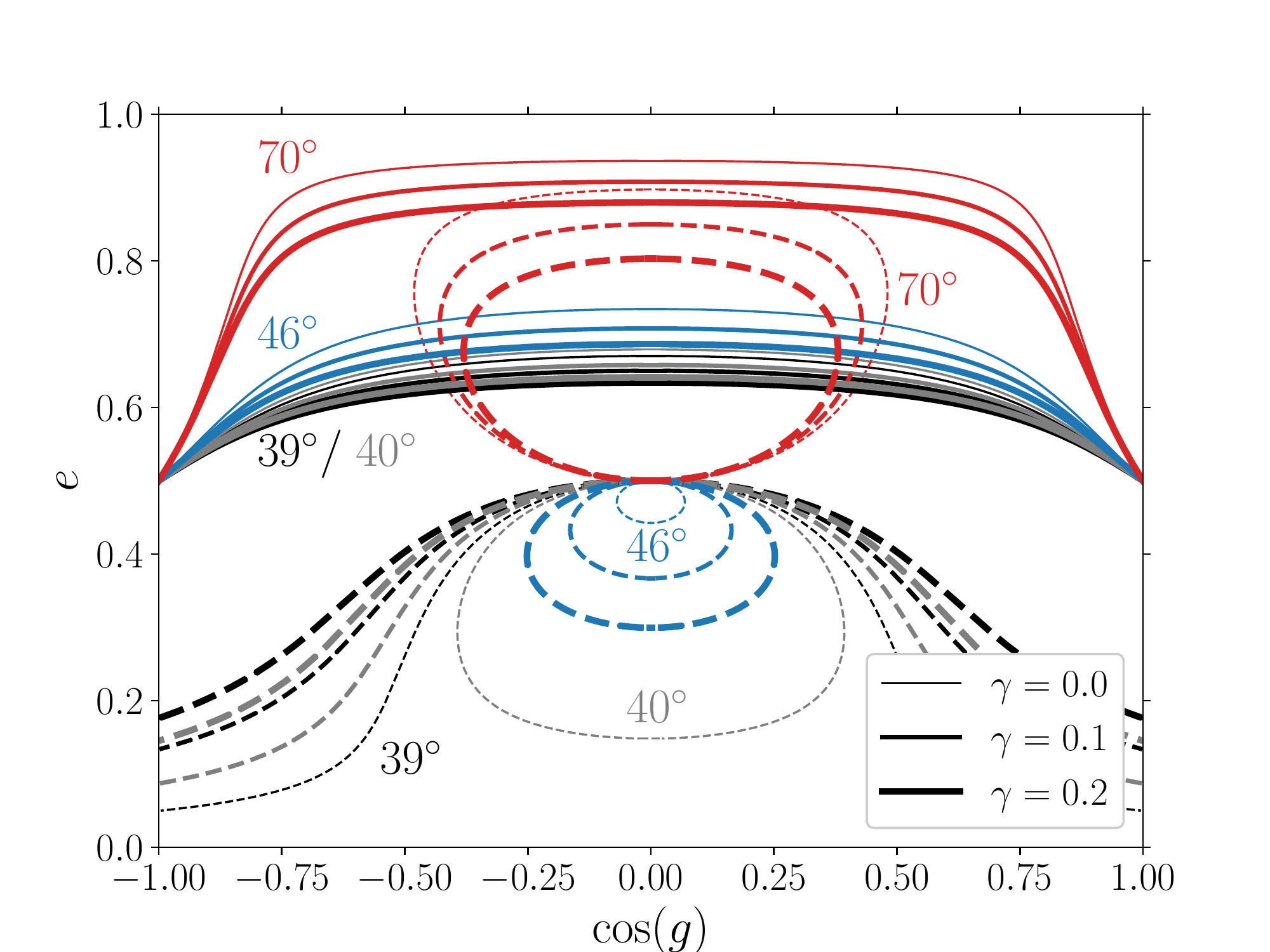}
\includegraphics[scale = 0.45, trim = 8mm 0mm 8mm 0mm]{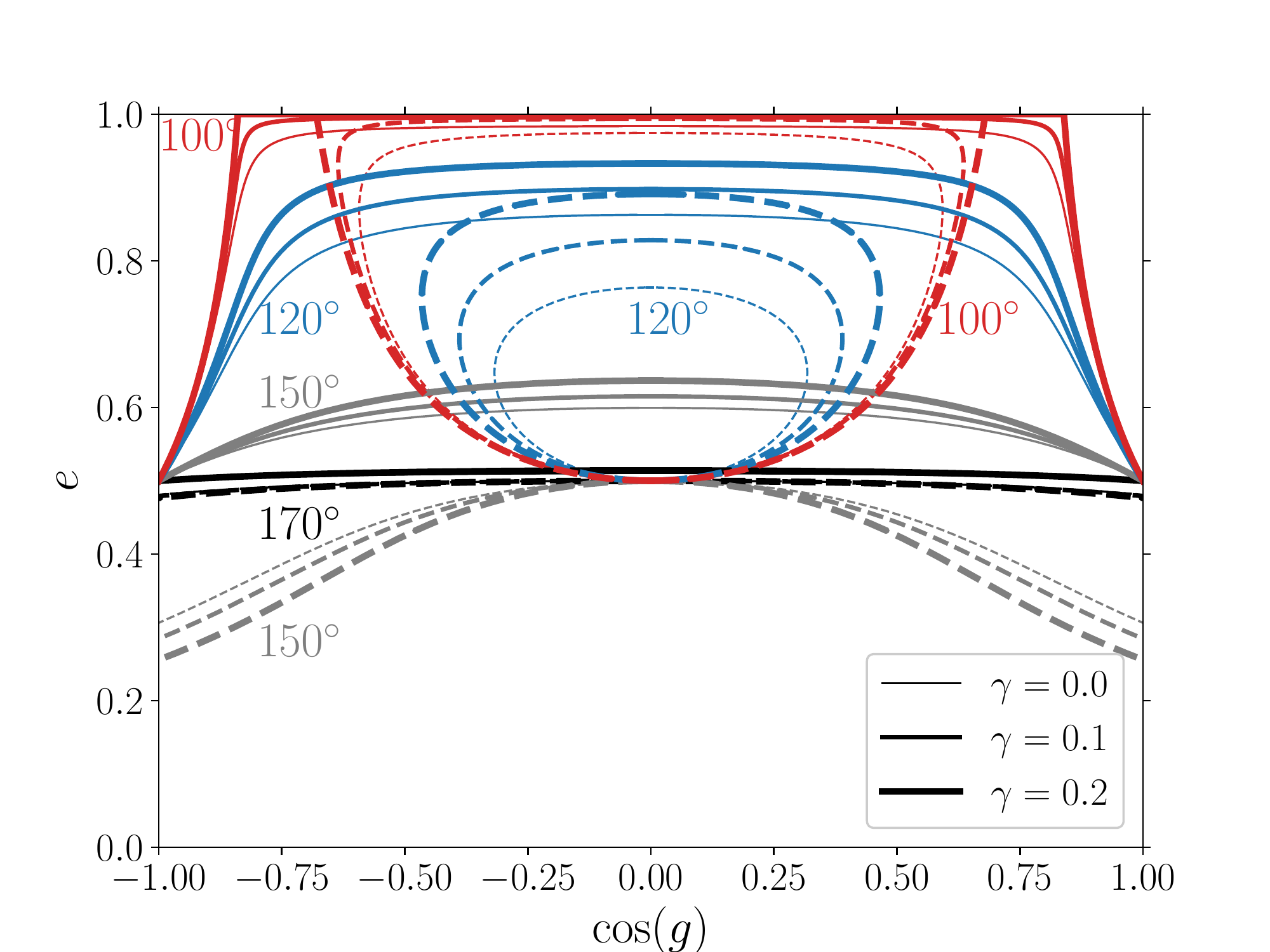}
\caption{Phase space plot of $(\cos g,e)$, where $g$ and $e$ are the inner orbit argument of periapsis and eccentricity, respectively, for a triple in the general case (at quadrupole order) with initial values $e_0=0.5$, $g_0=0$ (solid lines) or $g_0=\pi/2$ (dashed lines), and several values of the initial inclination (different colours). Thin, thicker, and thickest lines correspond to $\gamma=0$, 0.1, and 0.2, respectively, where $\gamma$ is defined in \Eq~(\ref{eq:gammadef}). In the top panel (prograde cases), we include solutions with $i_{\rel,\,0} = 39^\circ$ (black), $i_{\rel,\,0} = 40^\circ$ (grey), $i_{\rel,\,0} = 46^\circ$ (blue), and $i_{\rel,\,0} = 70^\circ$ (red). In the bottom panel (retrograde cases), included are $i_{\rel,\,0} = 170^\circ$ (black), $i_{\rel,\,0} = 150^\circ$ (grey), $i_{\rel,\,0} = 120^\circ$ (blue), and $i_{\rel,\,0} = 100^\circ$ (red). }
\label{fig:cont2}
\end{figure}

\subsection{Phase space plots}
\label{sect:gen:cont}
Using the same procedure as in \S~\ref{sect:tp:cont}, one can plot phase space trajectories in the $(\cos g,e)$ plane for $\gamma\neq 0$. This is done in the top panel of \F~\ref{fig:cont2}, where we repeat the curves for $\gamma=0$ from \F~\ref{fig:cont}, and include two additional sets with the same initial conditions, but with $\gamma=0.1$ and $\gamma=0.2$. 

In the top panel of \F~\ref{fig:cont2}, the initial orientations are prograde ($\theta_0>0$, or $0<i_{\rel,\,0}<90^\circ$). Comparing the different curves with increasing $\gamma$, one can see a shift toward smaller eccentricities. Increasing $\gamma$ means that the initial inclination is effectively reduced. However, this behaviour applies for prograde orbits. In the bottom panel of \F~\ref{fig:cont2}, we show phase-space curves for some retrograde cases ($\theta_0<0$, or $90<i_{\rel,\,0}<180^\circ$). Here, increasing $\gamma$ can decrease or increase the maximum eccentricities, depending on the initial relative inclination. We shall discuss this symmetry breaking and enhanced excitation as a function of $\gamma$ in more detail below in \S~\ref{sect:gamma}.

The shift due to non-zero $\gamma$ implies a different boundary between circulating and librating solutions. For example, comparing the curve with $g_0=\pi/2$, $i_{\rel,\,0}=40^\circ$, and $\gamma=0$ (thin dashed grey line) to the curves with $g_0=\pi/2$, $i_{\rel,\,0} =40^\circ$, and $\gamma>0$ (thicker dashed grey lines), one can observe a switch from librating to circulating solutions due to different $\gamma$. This can be quantified by setting $\cos 2g=\pm1$ with $e=0$, similarly as in \S~\ref{sect:tp:cont} but now not assuming $\gamma=0$, giving the condition $\CZ=0$ for the boundary between circulation and libration, where
\begin{align}
\label{eq:cz}
\CZ \equiv \frac{1}{12} \left [ c_0 - 6 \left ( \TZ-\gamma \right)^2 + 2 \right ].
\end{align}
As before, $\CZ>0$ for circulation, and $\CZ<0$ for libration, but $\CZ$ is now also a function of $\gamma$. Evidently, $\CZ(\gamma\rightarrow0) = \CZz$. With the definitions of $c_0$ and $\TZ$, $\CZ$ can be rewritten in terms of $e_0$, $\theta_0$, and $\gamma$ as
\begin{align}
\CZ = e_0^2 \left [1 - \frac{1}{2} e_0^2 \gamma^2 + \sqrt{1-e_0^2} \gamma \theta_0 - \frac{5}{2} \left(1-\theta_0^2 \right ) \sin^2 (g_0) \right ],
\end{align}
which reduces to \Eq~(\ref{eq:czz2}) as $\gamma \rightarrow 0$. Furthermore, setting $\CZ=0$, one can solve for $\gamma$ to find the critical value of $\gamma$, $\gamma_\crit$, for the system to lie exactly on the boundary between circulation and libration:
\begin{align}
\label{eq:gammacrit}
\gamma_\crit = \frac{1}{e_0^2} \left [ \sqrt{1-e_0^2} + \sqrt{\theta_0^2 + e_0^2 \left (2-5 \sin^2 g_0 - \theta_0^2 + 5 \theta_0^2 \sin^2 g_0 \right )} \right ].
\end{align}
Note that this relation does not apply if $e_0=0$; in the latter case, $\CZ=0$ so the system always lies on the boundary, irrespective of $\gamma$ (and $\theta_0$ and $g_0$).

To obtain the location of the fixed point, we again use \Eq~(\ref{eq:eomg}) and solve for $\theta_0$ with $\dot{g}=0$, and $g=\pi/2$ and $e=e_0$, but now we include the term $\propto 1/G_2$. This gives
\begin{align}
\label{eq:fpgen}
\theta_{0,\,\FP} = \sqrt{\frac{3}{5}\left (1-e_0^2 \right )} \left [ \sqrt{1 + \frac{\gamma^2}{15} \left (1+4e_0^2 \right )^2} - \frac{\gamma}{\sqrt{15}} \left(1+4 e_0^2 \right ) \right ].
\end{align}
As $\gamma\rightarrow 0$, \Eq~(\ref{eq:fpgen}) reduces to the test particle case, \Eq~(\ref{eq:fptp}). Generally, $\theta_{0,\,\FP}$ for $\gamma \neq0$ is reduced compared to the case $\gamma=0$.

\subsection{Eccentricity extrema}
\label{sect:gen:ecc}
\subsubsection{General case}
\label{sect:gen:ecc:gen}
Eccentricity extrema can be obtained from energy and angular momentum conservation and setting $g$ appropriately, as before in \S~\ref{sect:tp:ecc}. We remark, however, that analytic solutions are less straightforward to find in this case, since the relation between $\theta$ and $e$ (cf. \Eq~\ref{eq:theta}) is now more complicated. 

Nevertheless, we found closed-form analytic solutions which are comparatively compact. Let
\begin{subequations}
\begin{align}
A &\equiv 13 \gamma ^4-16 \gamma ^3 \TZ +48 \gamma  \TZ +\gamma ^2 \left(-24\CZ+4 \TZ ^2-6\right)+9; \\
\nonumber B &\equiv 35 \gamma ^6-264 \gamma ^5 \TZ +216 \gamma  \TZ -3 \gamma ^4 \left(60 \CZ-146 \TZ ^2+33\right) \\
& -16 \gamma ^3 \TZ  \left(18
   \CZ+13 \TZ ^2+9\right)-9 \gamma ^2 \left(12 \CZ-34 \TZ ^2+3\right)+27,
\end{align}
\end{subequations}
and
\begin{align}
\varphi \equiv \mathrm{arctan2} \left ( \sqrt{A^3 - B^2}, B \right ),
\end{align}
where $\mathrm{arctan2}(y,x)$ is the 2-argument arctangent function. The maximum eccentricity (in both circulating and librating cases) is then given by
\begin{align}
\label{eq:xmingen}
x_\min = \frac{1}{12\gamma^2} \left [ 3 + 5 \gamma^2 + 8 \gamma \TZ  - 2\sqrt{A} \sin \left (\frac{1}{3} \varphi + \frac{\pi}{6} \right ) \right ].
\end{align}
In the librating case ($\CZ<0$), the minimum eccentricity is given by
\begin{align}
\label{eq:xmaxgenlib}
x_\max = \frac{1}{12\gamma^2} \left [ 3 + 5 \gamma^2 + 8 \gamma \TZ  + 2\sqrt{A} \sin \left (\frac{1}{3} \varphi - \frac{\pi}{6} \right ) \right ], \quad (\CZ<0),
\end{align}
whereas in the circulating case ($\CZ>0$), the minimum eccentricity is given by
\begin{align}
\label{eq:xmaxgencirc}
\nonumber x_\max = \frac{1}{\gamma^2} \left [ 1 + \gamma \TZ - \sqrt{1 + 2 \gamma \TZ - 2 \gamma^2 + 2 \gamma^2 \CZ + \gamma^2 \left (\TZ - \gamma \right )^2} \right ], \\
\quad (\CZ>0).
\end{align}

The above expressions assume that $A^3 - B^2 \geq 0$. This is the case for nearly all numerical examples presented below in \S~\ref{sect:gamma}. For a small number of initial conditions, $A^3-B^2<0$. In these cases, only one real eccentricity stationary point exists, and is given by
\begin{align}
x_\min = \left \{ \begin{array}{ll}
\displaystyle \frac{1}{12\gamma^2} \left [ 3 + 5 \gamma^2 + 8 \gamma \TZ  - \frac{A}{a} - a \right ], & B + \sqrt{B^2-A^3} < 0; \\
\displaystyle \frac{1}{12\gamma^2} \left [ 3 + 5 \gamma^2 + 8 \gamma \TZ  + \frac{A}{b} + b \right ], & B + \sqrt{B^2-A^3} > 0, \\
\end{array} \right.
\end{align}
where
\begin{subequations}
\begin{align}
a &\equiv \left ( -B - \sqrt{B^2-A^3} \right )^{1/3}; \\
b &\equiv \left ( B + \sqrt{B^2-A^3} \right )^{1/3}. 
\end{align}
\end{subequations}

From these general expressions, several limiting cases can be considered, as shown below.

\subsubsection{Limiting cases}
\label{sect:gen:ecc:lim}
It can be shown that, when the limit $\gamma \rightarrow 0$ is taken, \Eqs~(\ref{eq:xmingen}), (\ref{eq:xmaxgenlib}), and (\ref{eq:xmaxgencirc}) reduce to the test particle equivalent \Eqs~(\ref{eq:xmintp}), (\ref{eq:xmaxtplib}), and (\ref{eq:xmaxtpcirc}), respectively. 

Another simplification is to set $e_0=0$ (but with $\gamma \neq 0$; this limit was considered in Appendix A of \citealt{2013MNRAS.431.2155N}). In this case, the dependence on $g_0$ disappears and $x_\max=1$, whereas
\begin{align}
\label{eq:xminzeroe0}
\nonumber x_\min &= \frac{1}{8\gamma^2} \Biggl [ 3 + 9 \gamma^2 + 8 \gamma \theta_0 \\
&\qquad  - \sqrt{9 + \gamma \left ( 48 \theta_0 + \gamma \left (54 + \gamma^2 - 16 \gamma \theta_0 -16 \theta_0^2 \right ) \right )} \Biggl ].
\end{align}
From \Eq~(\ref{eq:xminzeroe0}), it is straightforward to show that $\mathrm{d}x_\min/\mathrm{d} \theta_0 = 0$ if $\theta_0 = \theta_{0,\,\peak}$ ($-1\leq \theta_0\leq 1$), where
\begin{align}
\label{eq:theta0peak}
\theta_{0,\,\peak} = -\gamma.
\end{align}
In other words, the maximum eccentricity as a function of $\theta_0$ peaks at $\theta_0=-\gamma$. This reduces to $\theta_{0,\,\peak} = 0$ if $\gamma =0$. Therefore, in the test particle limit, there is complete symmetry in the eccentricity behaviour around $\theta_0=0$. In the general case, the symmetry is broken, and $e_\max$ becomes centered around $\theta_0 = -\gamma$. We note that this result was mentioned before by \citet{2015MNRAS.447..747L}.

In addition, by setting $\mathrm{d}x_\min/\mathrm{d} \gamma = 0$ with $x_\min$ given by \Eq~(\ref{eq:xminzeroe0}), we can find the value of $\gamma$ for which $e_\max$ shows a local maximum if $e_0=0$,
\begin{align}
\label{eq:gammamax}
\gamma_\max = -\theta_0.
\end{align}
We will consider these limiting cases quantitatively in \S~\ref{sect:gamma}.

\subsection{Timescales}
\label{sect:gen:time}
\Eq~(\ref{eq:tzlk}) for the ZLK eccentricity timescale still applies in the non-test-particle limit when \Eq~(\ref{eq:theta}) is used to express $\theta$ as a function of $x$, with $\gamma \neq 0$. However, we found the resulting integral to be intricate to solve in closed analytic form. Therefore, we will proceed by solving \Eq~(\ref{eq:tzlk}) for the case $\gamma \neq 0$ numerically (but with the limits calculated using the analytic relations in \S~\ref{sect:gen:ecc}).

\section{Dependences on $\gamma$}
\label{sect:gamma}
In this section, we explore in detail the properties of ZLK oscillations in the non-test-particle limit using the (semi)analytic expressions presented in \S~\ref{sect:gen}, as well as using results from numerically integrating the equations of motion. The latter is achieved by using the \textsc{SecularMultiple} code \citep{2016MNRAS.459.2827H,2018MNRAS.476.4139H,2020MNRAS.494.5492H}, which is freely available\footnote{\href{https://github.com/hamers/secularmultiple}{https://github.com/hamers/secularmultiple}}. This code is based on an expansion of the Hamiltonian of the system in terms of ratios of separations of orbits (it is not necessarily restricted to hierarchical triples, but we apply it here to triples). The Hamiltonian is subsequently orbit averaged, and the resulting equations of motion (in the form of a set of ordinary differential equations) is solved numerically.

For the purposes of this section, we developed a \textsc{Python} script, \texttt{ZLK.py}\footnote{This script is freely available at \href{https://github.com/hamers/ZLK}{https://github.com/hamers/ZLK}.},  which implements the (semi)analytic expressions of \S~\ref{sect:gen} (fast evaluation), but also (optionally) allows to integrate equivalent triple systems numerically using \textsc{SecularMultiple} (more computationally expensive). 

To set up the numerical integrations in \textsc{SecularMultiple}, we fix all parameters of the triple except for $a_2$. For a given $\gamma$, we then compute the corresponding value of $a_2$ using \Eq~(\ref{eq:gammadef}). The system is then integrated for a duration of $10 \, L_1/C_2$ (cf. \Eq~\ref{eq:l1divc2}), with the inclusion of the quadrupole-order expansion terms only (see \S~\ref{sect:discussion:oct} for a discussion on the importance of octupole-order terms). During the integration, the code stops at eccentricity extrema, which are recorded and used to return the minimum and maximum eccentricities, as well as the timescale of the eccentricity oscillations. 

Here, we focus on the maximum eccentricities (\S~\ref{sect:gamma:ecc}), as well as the eccentricity oscillation timescales (\S~\ref{sect:gamma:t}). We briefly discuss conditions for orbital flips in \S~\ref{sect:gamma:flip}.

\begin{figure}
\center
\includegraphics[scale = 0.45, trim = 8mm 0mm 8mm 0mm]{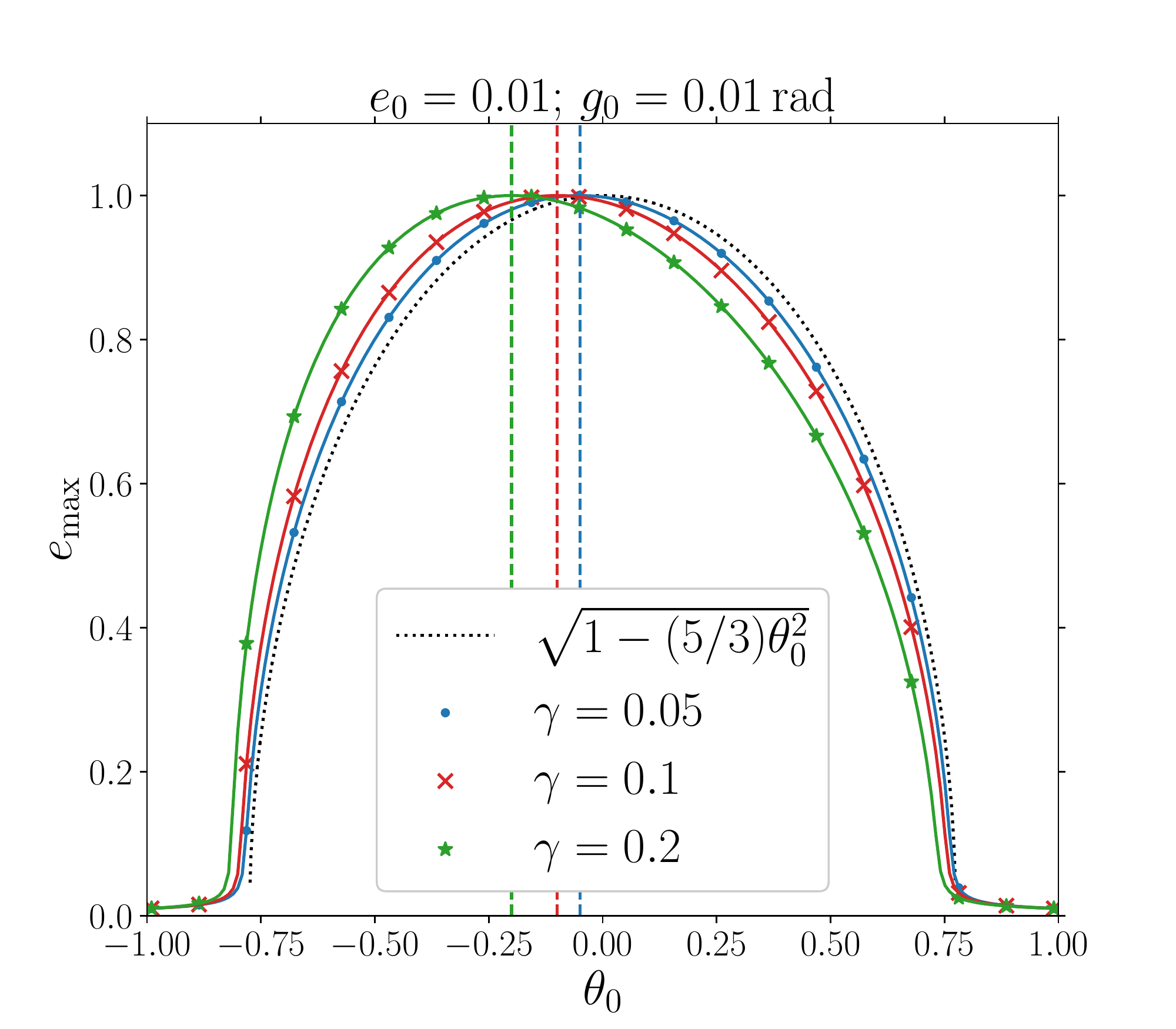}
\includegraphics[scale = 0.45, trim = 8mm 0mm 8mm 0mm]{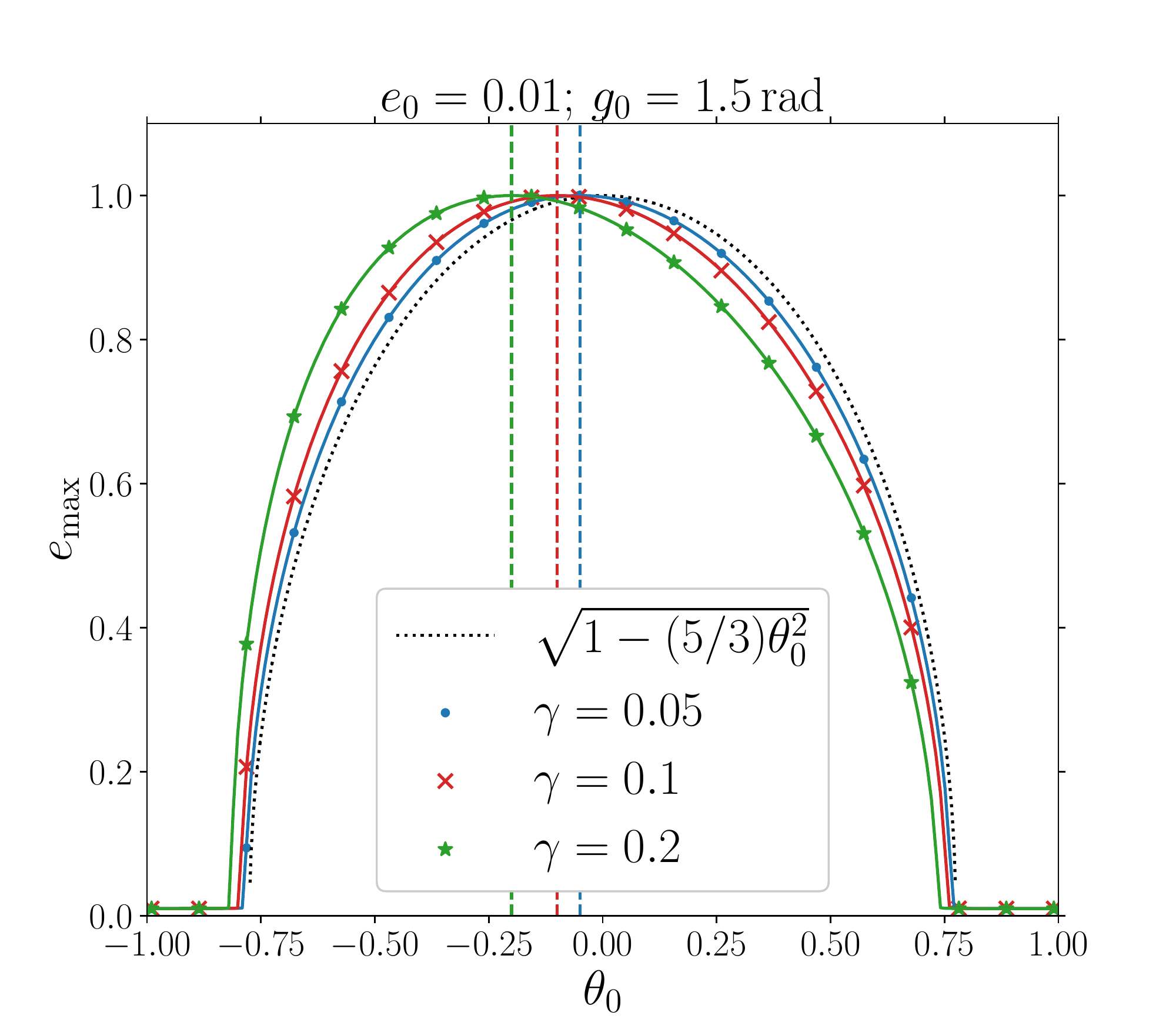}
\caption{Maximum eccentricities as a function of the cosine of the initial relative inclination, $\theta_0 \equiv \cos (i_{\rel,\,0})$. Solid lines: analytic results using the expressions of \S~\ref{sect:gen:ecc}. Different colours correspond to different values of $\gamma$, indicated in the legend. Markers indicate results from numerical integrations of the equations of motion. The top and bottom panels correspond to $g_0=0.01$ rad and $g_0=1.5$ rad, respectively, whereas $e_0=0.01$ in both panels. The black dotted lines show the `classical' result (\Eq~\ref{eq:xmincan}) which applies in the test particle limit with $e_0=0$. The coloured vertical dashed lines show $\theta_0=-\gamma$, which is the value for which $e_\max$ peaks as a function of $\theta_0$ assuming $e_0=0$ (cf. \Eq~\ref{eq:theta0peak}). }
\label{fig:emax1set1}
\end{figure}

\begin{figure}
\center
\includegraphics[scale = 0.45, trim = 8mm 0mm 8mm 0mm]{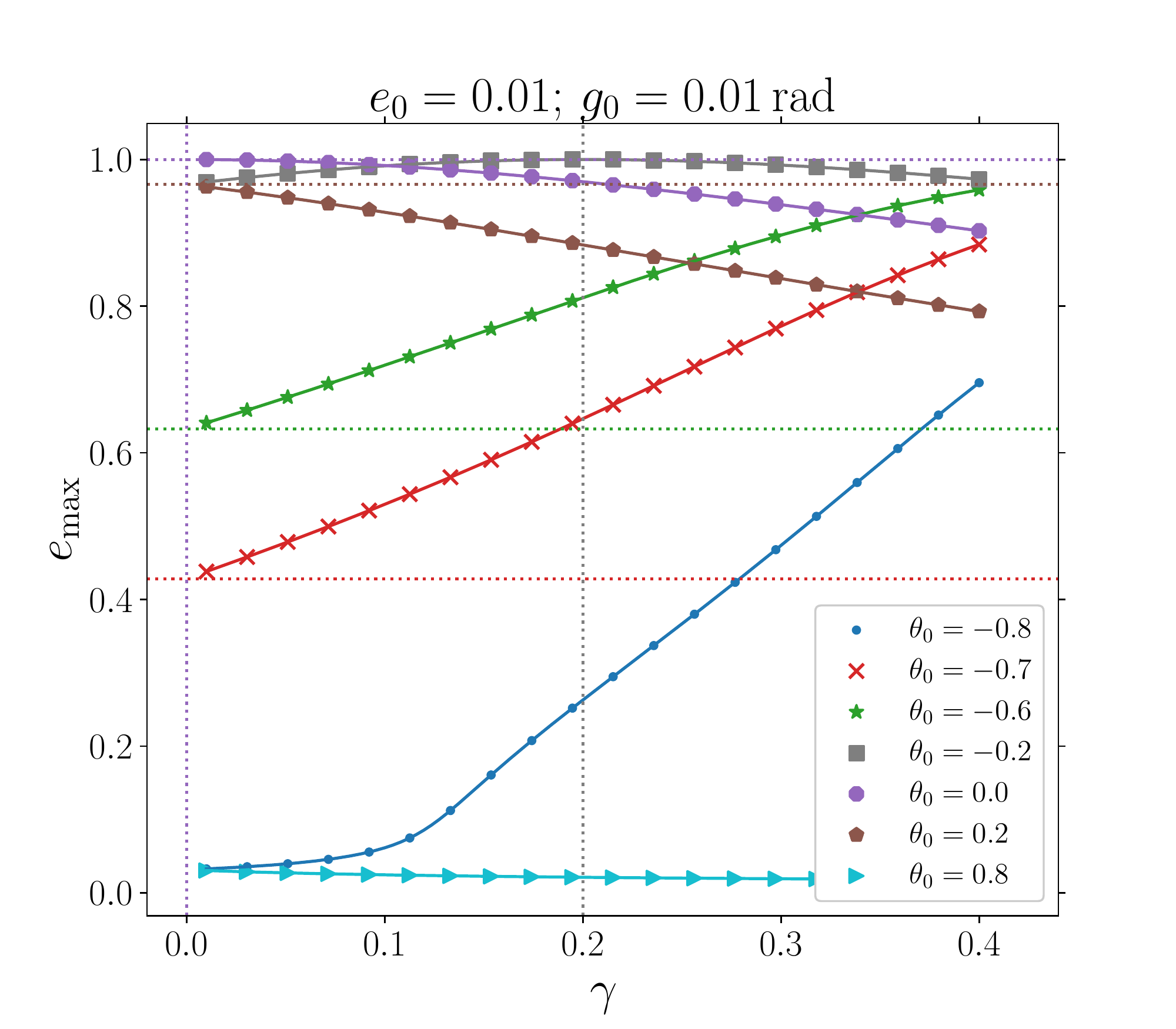}
\includegraphics[scale = 0.45, trim = 8mm 0mm 8mm 0mm]{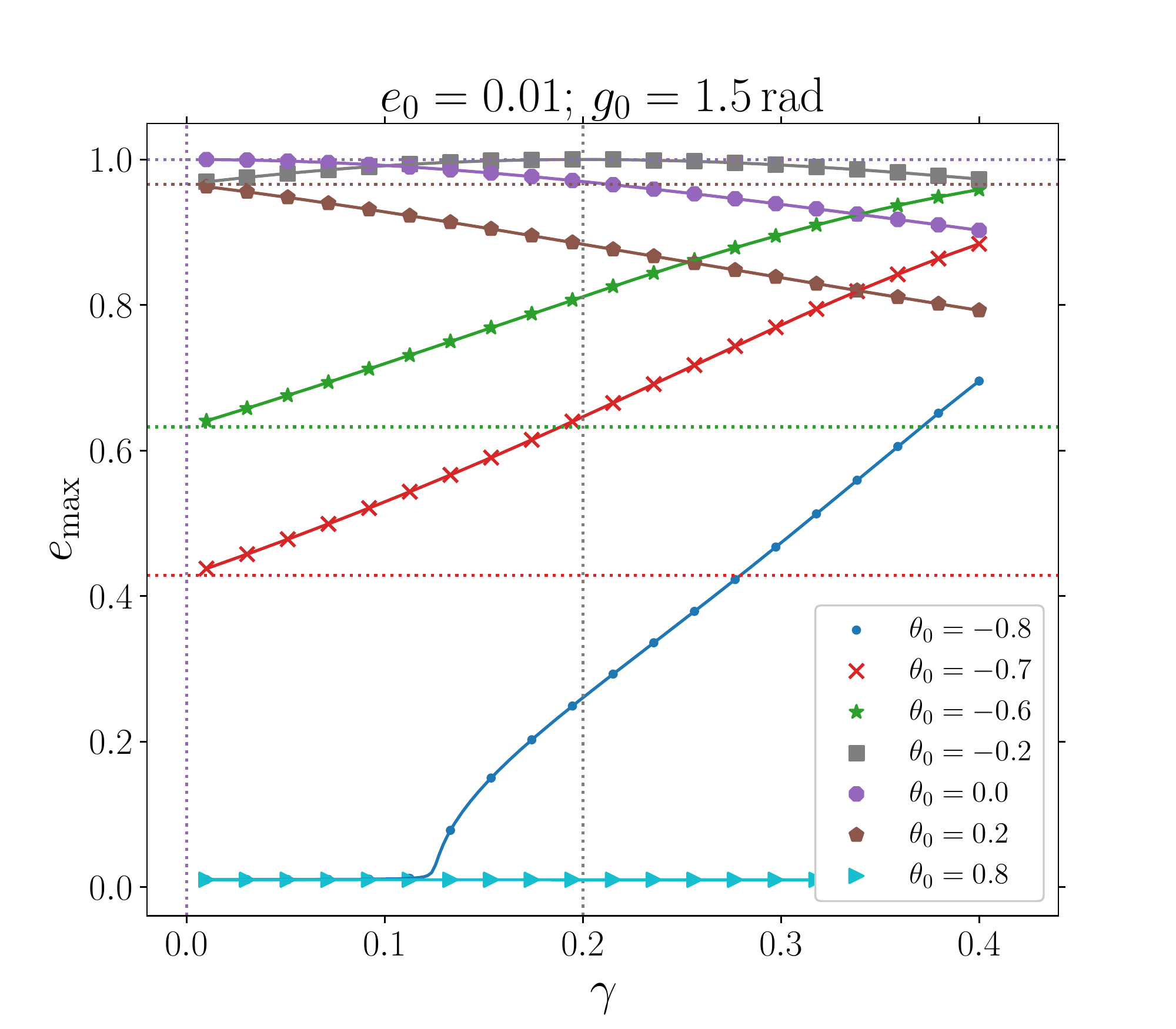}
\caption{Maximum eccentricities as a function of $\gamma$ for fixed $e_0$ and $g_0$, and several values of $\theta_0$. Solid lines: analytic results using the expressions of \S~\ref{sect:gen:ecc}. Different colours correspond to different values of $\theta_0$, indicated in the legend. Markers indicate results from numerical integrations of the equations of motion. The top and bottom panels correspond to $g_0=0.01$ rad and $g_0=1.5$ rad, respectively, whereas $e_0=0.01$ in both panels. The coloured horizontal dotted lines show the `classical' result (\Eq~\ref{eq:xmincan}) which applies in the test particle limit with $e_0=0$.}
\label{fig:emax2set1}
\end{figure}

\begin{figure}
\center
\includegraphics[scale = 0.45, trim = 8mm 0mm 8mm 0mm]{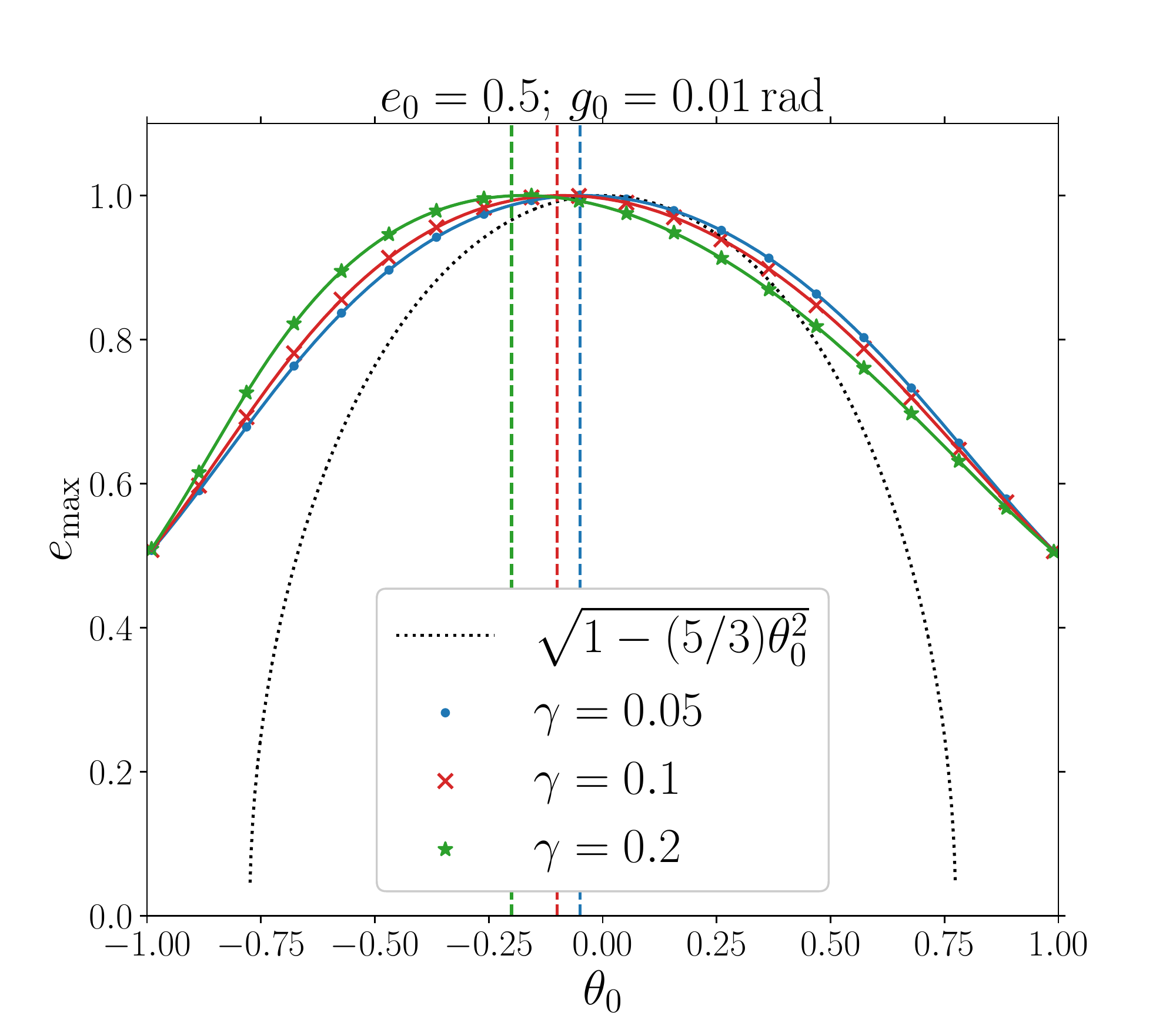}
\includegraphics[scale = 0.45, trim = 8mm 0mm 8mm 0mm]{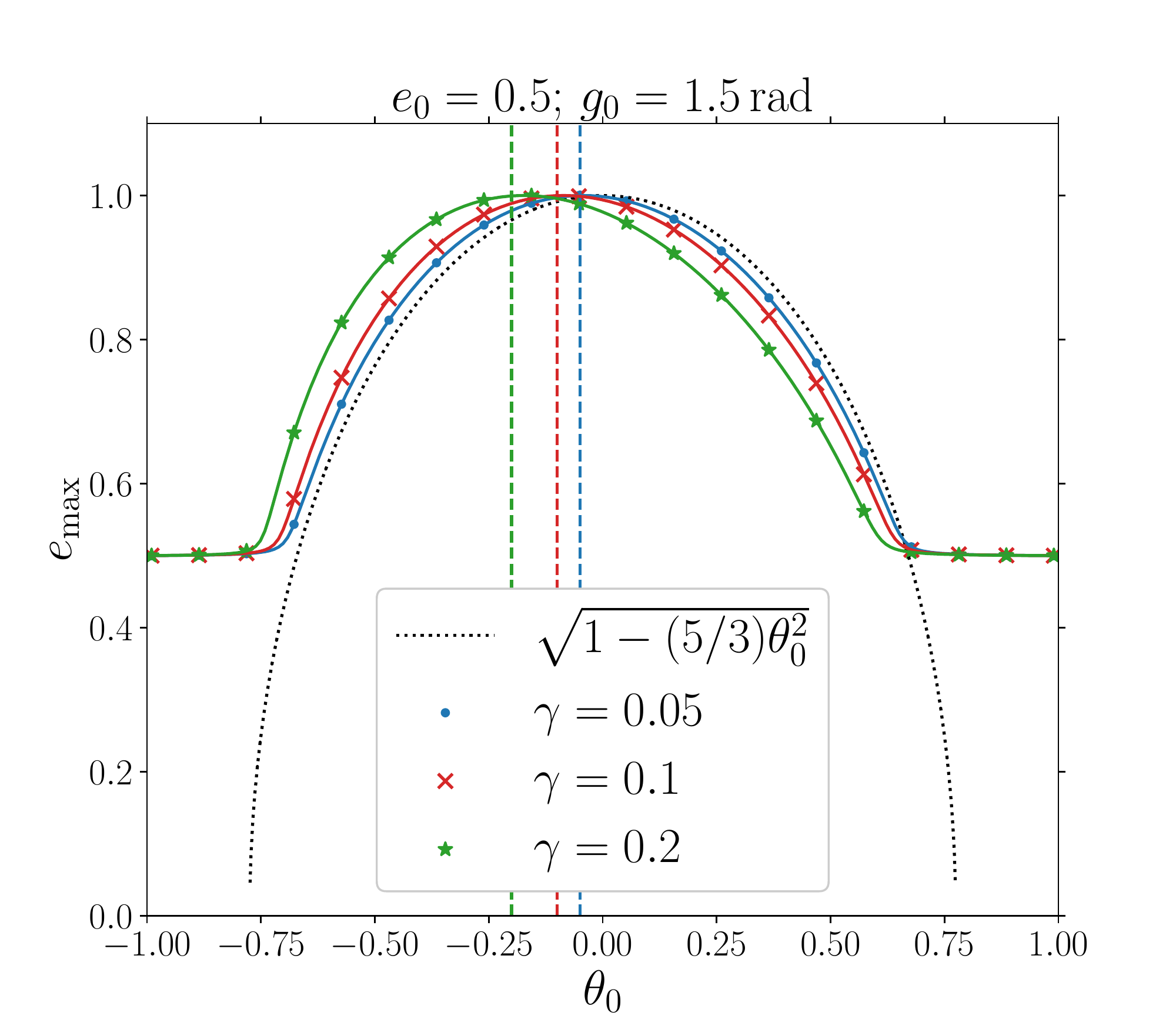}
\caption{Similar to \F~\ref{fig:emax1set1}, now with a higher initial eccentricity, $e_0=0.5$. }
\label{fig:emax1set2}
\end{figure}

\begin{figure}
\center
\includegraphics[scale = 0.45, trim = 8mm 0mm 8mm 0mm]{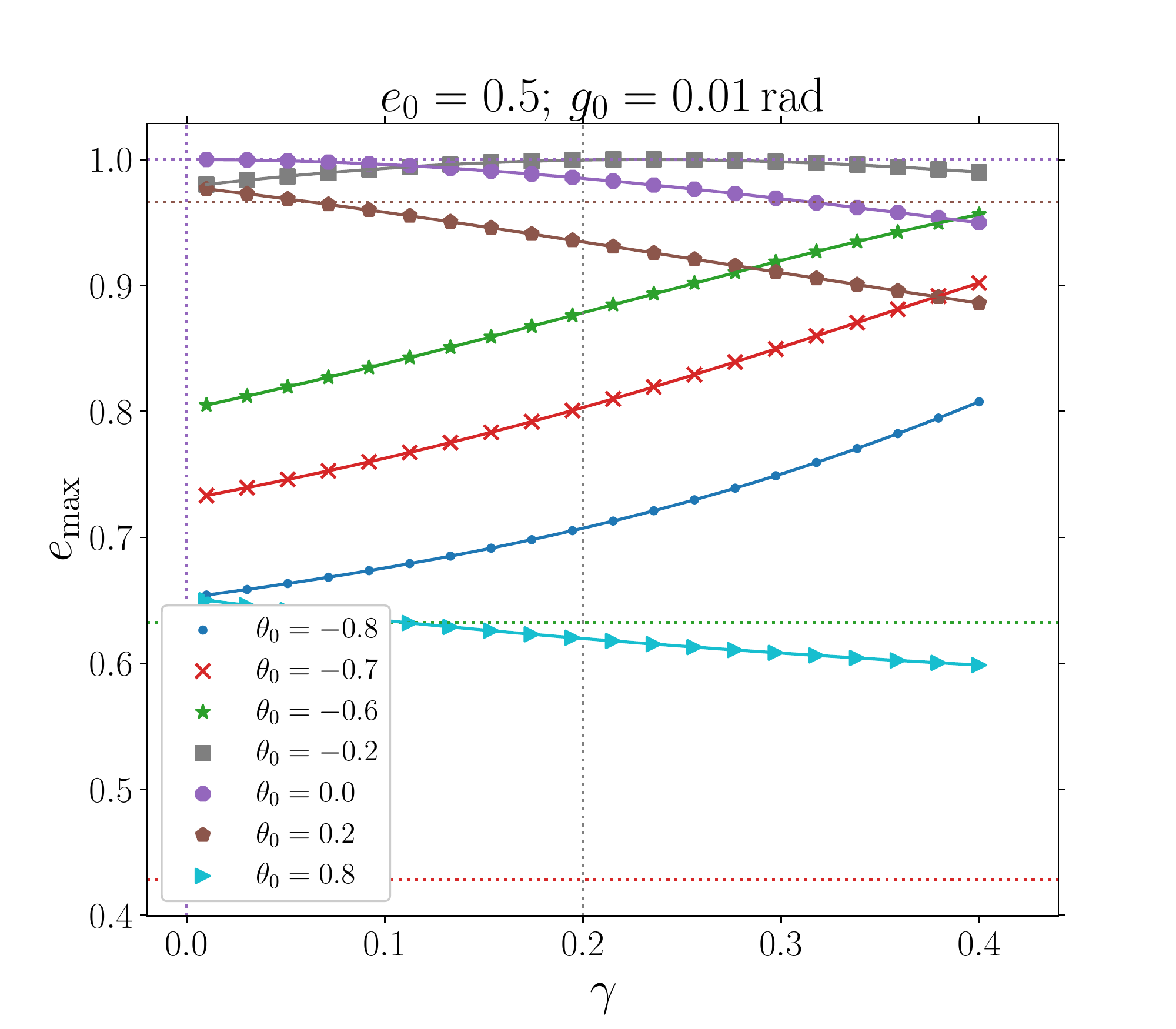}
\includegraphics[scale = 0.45, trim = 8mm 0mm 8mm 0mm]{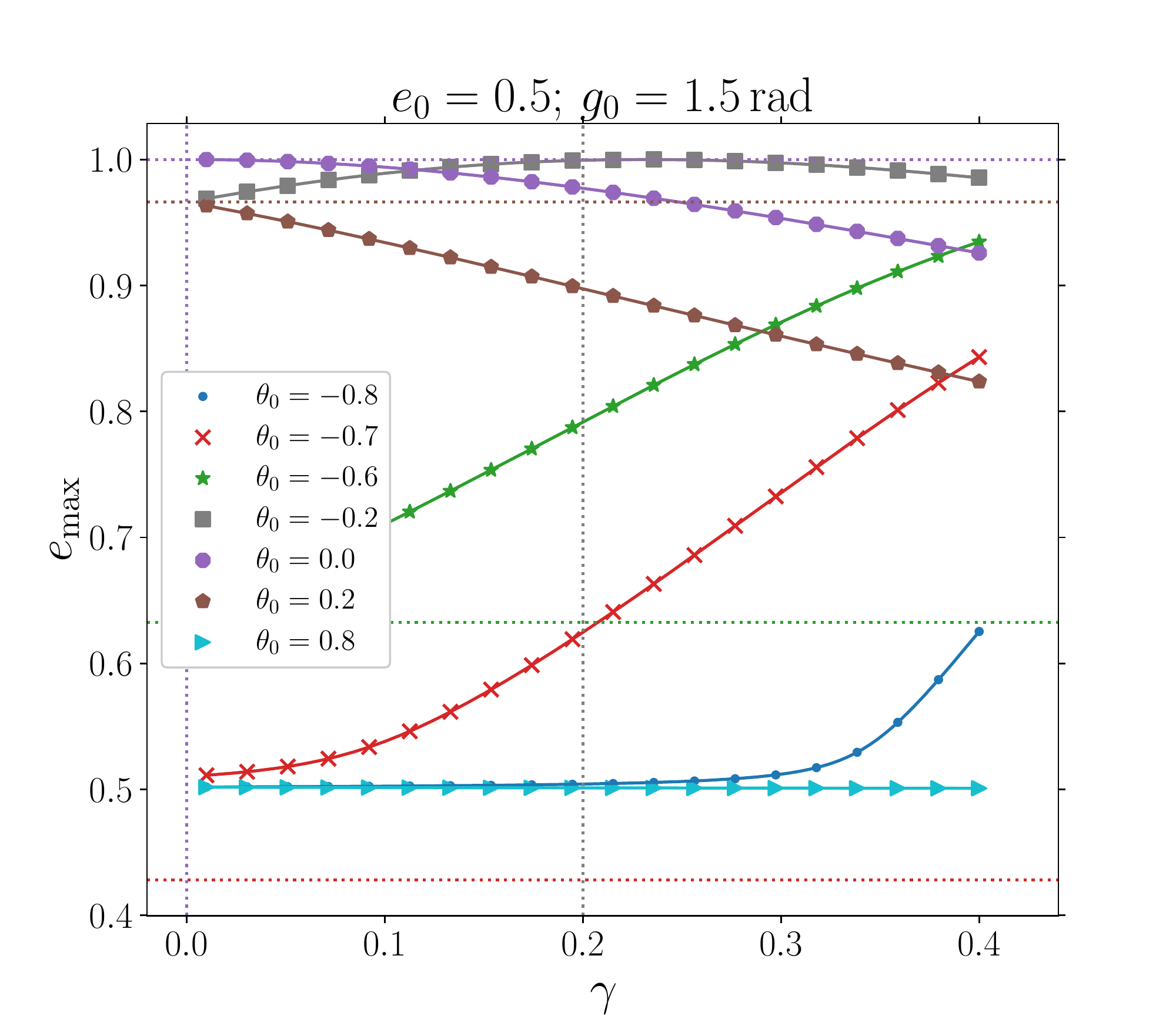}
\caption{Similar to \F~\ref{fig:emax2set1}, now with a higher initial eccentricity, $e_0=0.5$. }
\label{fig:emax2set2}
\end{figure}

\subsection{Maximum eccentricities}
\label{sect:gamma:ecc}
In \Fs~\ref{fig:emax1set1} through \ref{fig:emax2set2}, we show maximum eccentricities as a function of either $\theta_0$ or $\gamma$, fixing other parameters. Solid lines show analytic results from \S~\ref{sect:gen:ecc}, whereas markers indicate numerical results obtained with \textsc{SecularMultiple}. There are no discernible discrepancies between the analytic and numerical results. 

In \F~\ref{fig:emax1set1}, $e_0=0.01$ is close to zero; for small gamma, the test particle result with $e_0$ applies (cf. the black dotted lines). As $\gamma$ is increased, the maximum eccentricity curves shift to the left toward negative $\theta_0$, i.e., $e_\max$ peaks for retrograde orientations. The vertical coloured dashed lines show $\theta_0=\theta_{0,\,\peak}=-\gamma$, the analytic value for the peak in the limit that $e_0=0$ (cf. \Eq~\ref{eq:theta0peak}), which matches well with the solid lines/markers in this case. 

\F~\ref{fig:emax2set1} shows, for the same choices of $e_0$ and $g_0$, $e_\max$ as a function of $\gamma$ for several values of $\theta_0$. For prograde orientations ($\theta_0>0$), $e_\max$ decreases with increasing $\gamma$. However, for highly retrograde orientations, $e_\max$ increases with increasing $\gamma$. When the orbit is initially slightly retrograde (e.g., see the curves with $\theta_0=-0.2$), $e_\max$ initially increases but then decreases with increasing $\gamma$, i.e., there exists a local maximum as a function of $\gamma$. 

We interpret this behaviour as the result of two competing effects:
\begin{enumerate}
\item Increasing $\gamma$ implies a larger inner orbit angular momentum. The inner orbit therefore becomes more difficult for the outer orbit to torque, which reduces the efficiency of the excitation of its eccentricity. 
\item When the orbit is highly retrograde and the inner orbital angular momentum is sufficiently large, the inner orbit can torque the outer orbit, affecting the relative inclination. This can induce an effective inclination which is closer to $90^\circ$, implying more efficient eccentricity excitation (see \Eq~\ref{eq:xmincan}). 
\end{enumerate}

\begin{figure}
\center
\includegraphics[scale = 0.45, trim = 8mm 0mm 8mm 0mm]{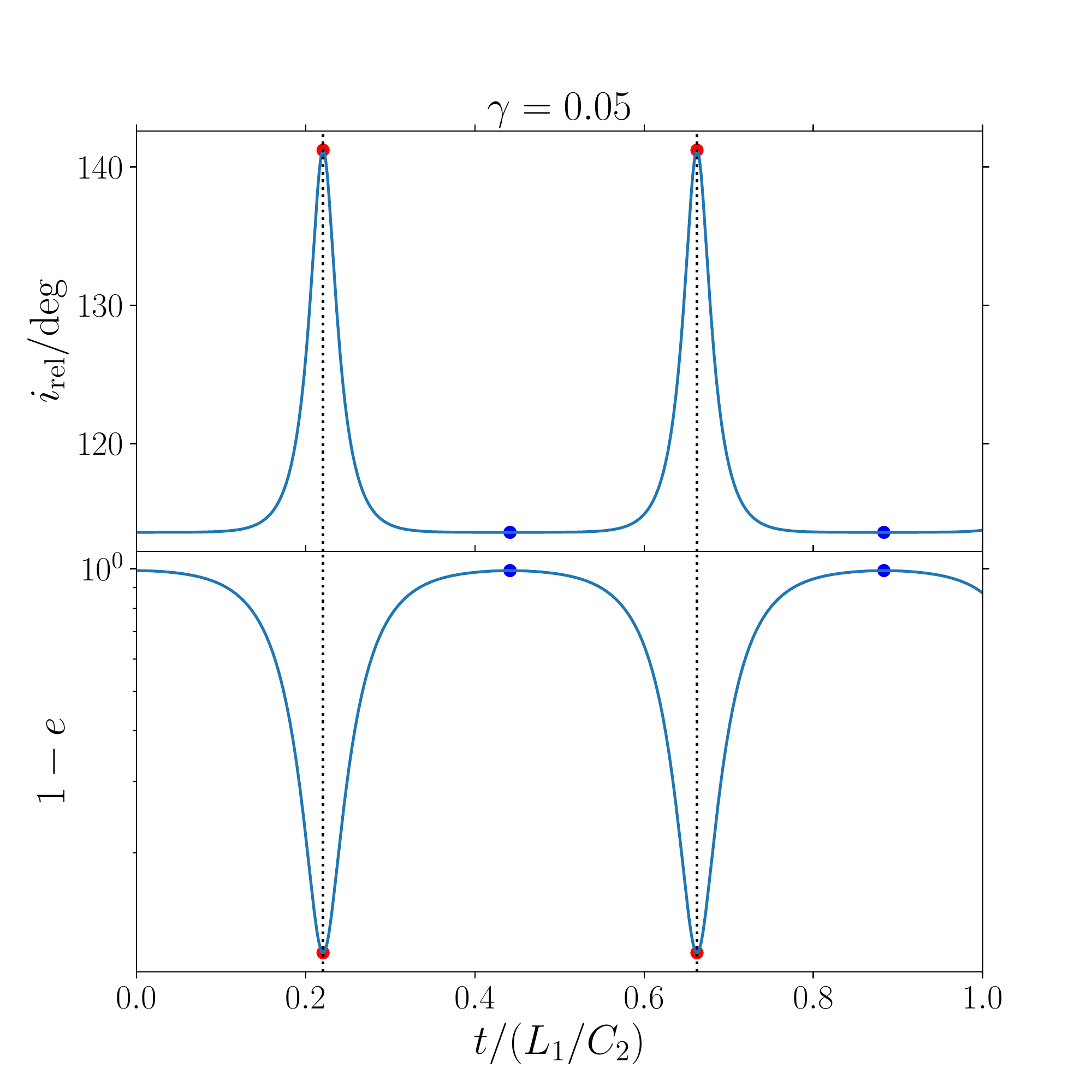}
\includegraphics[scale = 0.45, trim = 8mm 0mm 8mm 0mm]{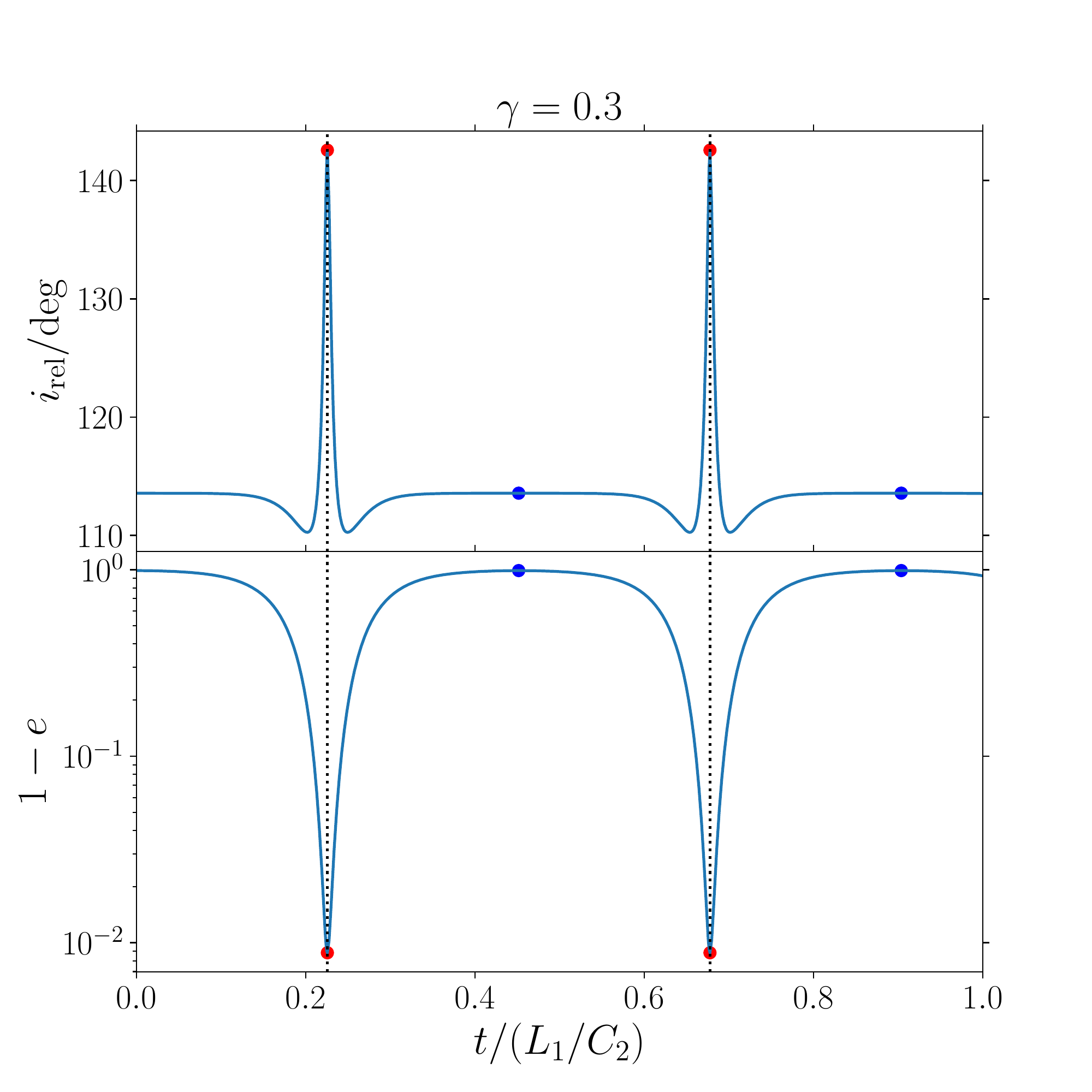}
\caption{Time evolution (from integration of the equations of motion) of the relative inclination (top panels) and the inner orbit eccentricity (bottom panels) for two values of $\gamma$: 0.05 (top set of panels), and 0.3 (bottom set of panels). The initial orientation is retrograde, with $\theta_0 = -0.4$. Other initial parameters are $e_0=0.01$, and $g_0 = 0.01\, \mathrm{rad}$. Blue and red dots indicate locations corresponding to eccentricity minima and maxima, respectively (determined via numerical root finding). Black vertical dotted lines also indicate eccentricity maxima. Time is normalised to $L_1/C_2$ (cf. \Eq~\ref{eq:l1divc2}). }
\label{fig:time}
\end{figure}

To illustrate the second effect, we show in \F~\ref{fig:time} an example of the time evolution (from integration of the equations of motion) of the relative inclination (top panels) and the inner orbit eccentricity (bottom panels) for two values of $\gamma$: 0.05 (top set of panels), and 0.3 (bottom set of panels). The initial orientation is retrograde, with $\theta_0 = -0.4$; other initial parameters are $e_0=0.01$, and $g_0 = 0.01\,\mathrm{rad}$. In the case of small $\gamma$, as $e$ increases, the relative inclination monotonically increases from the initial value to the value corresponding to eccentricity maximum. The term $\propto \TZ$ in \Eq~(\ref{eq:theta}) dominates in this case, and $\theta$, initially negative, decreases and becomes more negative as $e$ approaches its maximum value. When $\gamma=0.3$, the second term $\gamma (1-e^2)$ in \Eq~(\ref{eq:theta}) becomes important, and causes an initial increase in $\theta$ (such that $i_\rel$ gets closer to $90^\circ$). As $e$ increases further, the term $\gamma (1-e^2)$ starts to decrease in importance, and $\theta$ again decreases. This behaviour implies a relative inclination which is effectively closer to $90^\circ$, giving a significantly higher maximum eccentricity in this case. 

The competing nature of these effects explains why, for slightly retrograde orientations, there exists a local maximum in $e_\max$ as a function of $\gamma$. As discussed in \S~\ref{sect:gen:ecc}, when $e_0=0$, the location of this maximum as a function of $\gamma$ is given by $\gamma=\gamma_\max = - \theta_0$ (cf. \Eq~\ref{eq:gammamax}). Since always $\gamma \geq 0$, this implies that a local maximum can only occur for retrograde orbits. When $\theta_0=-0.2$, $\gamma_\max=0.2$, consistent with the location of the local maximum in \F~\ref{fig:emax1set1}.

\Fs~\ref{fig:emax1set2} and \ref{fig:emax2set2} show similar figures as \Fs~\ref{fig:emax1set1} and \ref{fig:emax2set1}, but now with $e_0=0.5$. Regarding the dependence on $\theta_0$, changing $e_0$ and $g_0$ mainly affects the edge regions of small and large $\theta_0$. Consequently, the locations of the eccentricity peaks are not much affected, and \Eq~(\ref{eq:theta0peak}) still gives a good approximation of the location of the maximum eccentricity peak with respect to $\theta_0$. 

No qualitative changes occur in the $(e_\max,\gamma)$ plots when increasing $e_0$ from 0.01 to 0.5. The maximum eccentricity still decreases with increasing $\gamma$ for prograde orientations; for highly retrograde orientations, $e_\max$ increases with increasing $\gamma$. For slightly retrograde orientations, there is a local maximum near $\gamma_\max=-\theta_0$.

\begin{figure}
\center
\includegraphics[scale = 0.45, trim = 8mm 0mm 8mm 0mm]{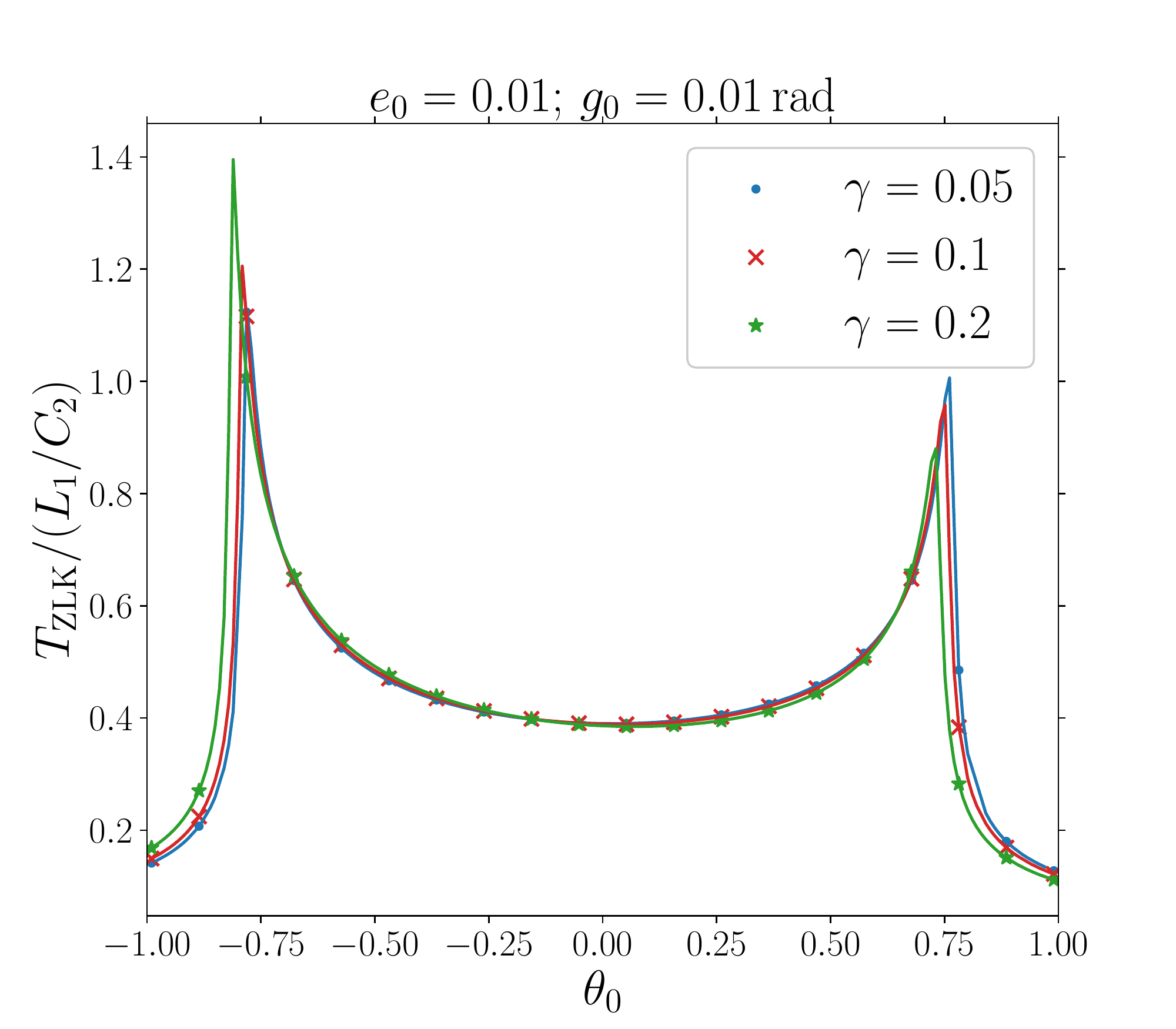}
\includegraphics[scale = 0.45, trim = 8mm 0mm 8mm 0mm]{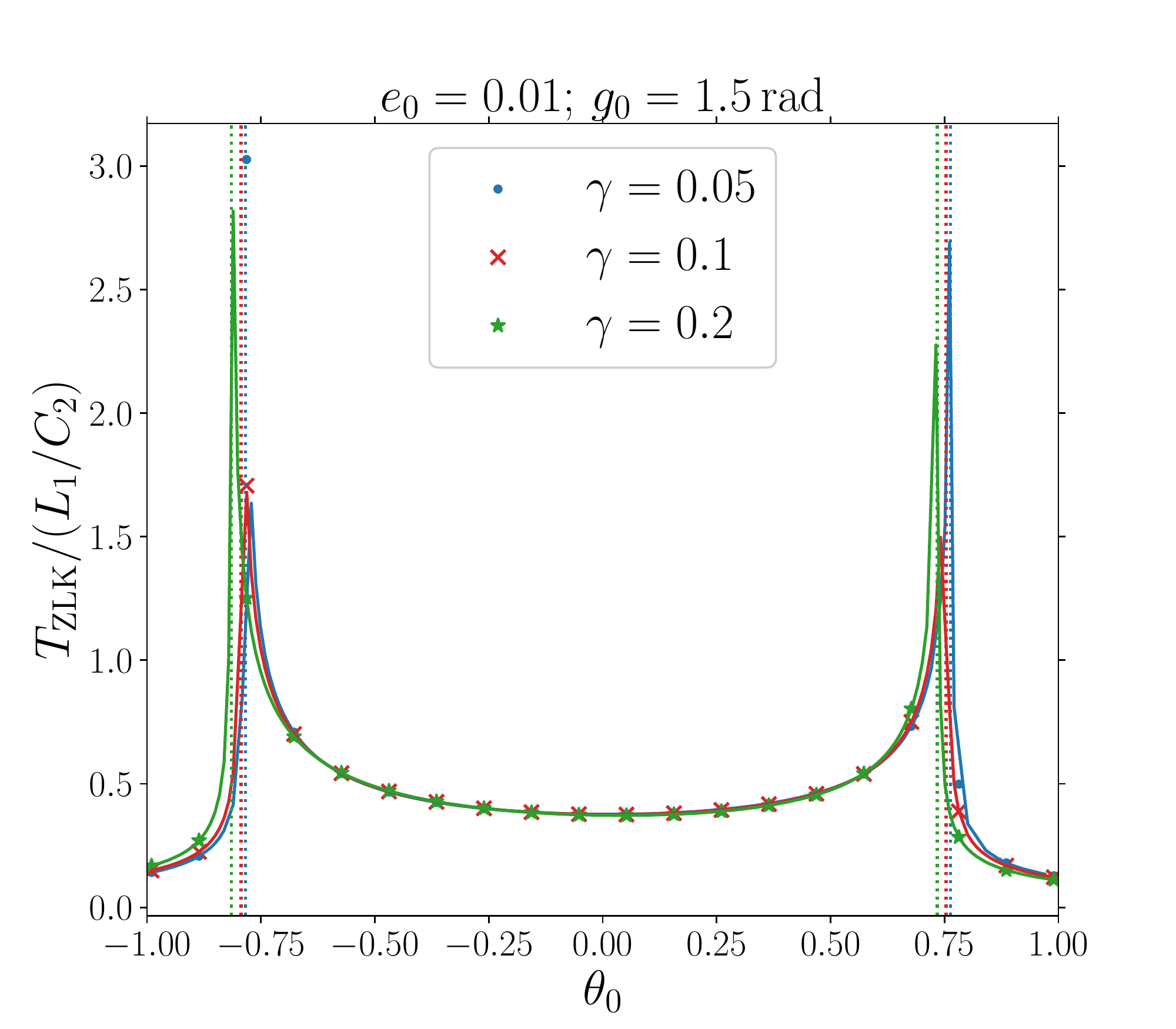}
\caption{Timescales of eccentricity oscillations as a function of the cosine of the initial relative inclination, $\theta_0 \equiv \cos (i_{\rel,\,0})$. The timescales are normalised to the quantity $L_1/C_2$ (cf. \Eq~\ref{eq:l1divc2}). Solid lines: numerical calculations of the integral \Eq~(\ref{eq:tzlk}). Different colours correspond to different values of $\gamma$, indicated in the legend. Markers indicate results from numerical integrations of the equations of motion. The top and bottom panels correspond to $g_0=0.01$ rad and $g_0=1.5$ rad, respectively, whereas $e_0=0.01$ in both panels. The vertical coloured dotted lines show the critical value of $\theta_0$ for the boundary between circulation and libration, which can be calculated analytically from \Eq~(\ref{eq:cz}).}
\label{fig:t1set1}
\end{figure}

\begin{figure}
\center
\includegraphics[scale = 0.45, trim = 8mm 0mm 8mm 0mm]{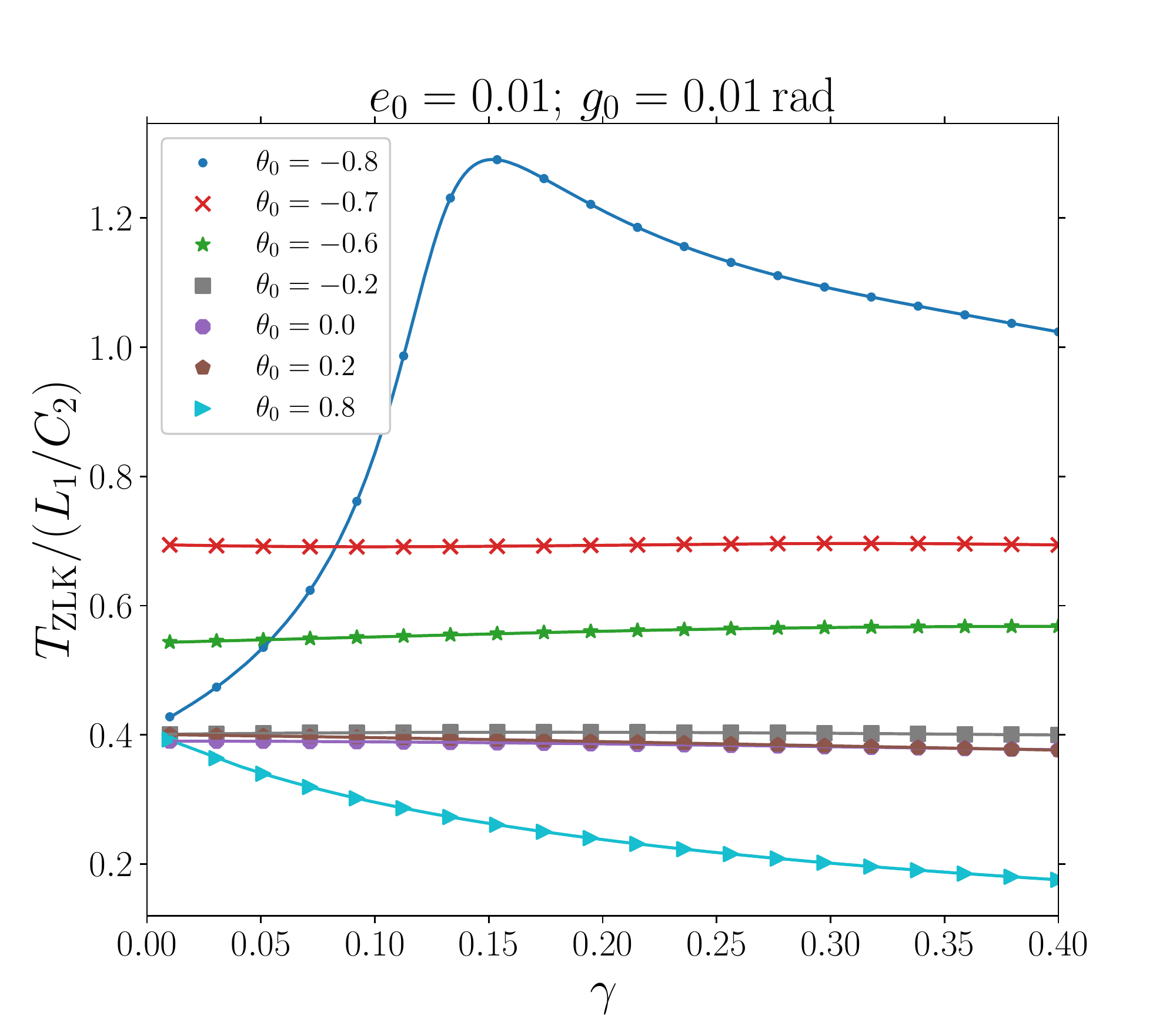}
\includegraphics[scale = 0.45, trim = 8mm 0mm 8mm 0mm]{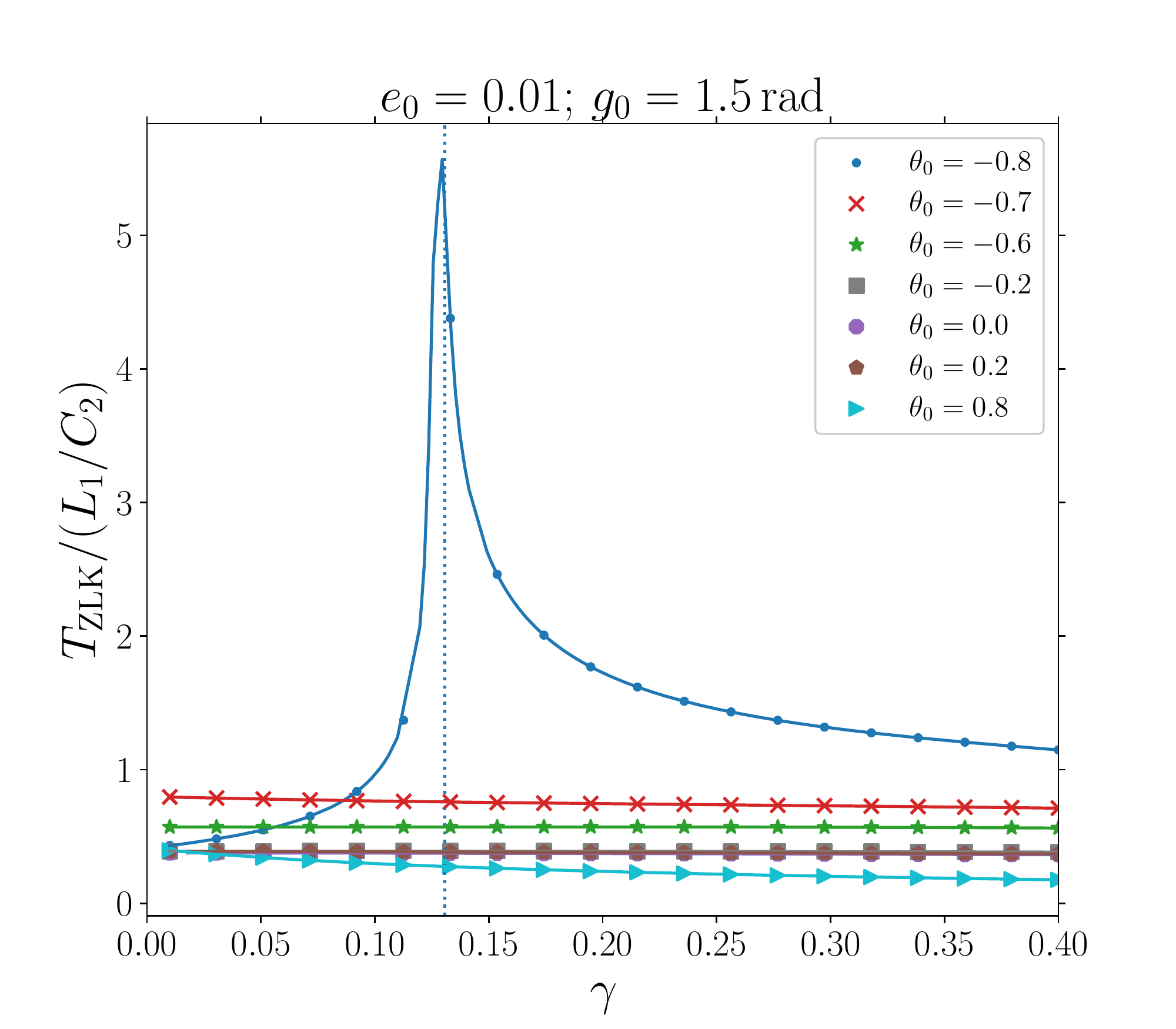}
\caption{Timescales of eccentricity oscillations as a function of $\gamma$ for fixed $e_0$ and $g_0$, and several values of $\theta_0$. The timescales are normalised to the quantity $L_1/C_2$ (cf. \Eq~\ref{eq:l1divc2}). Solid lines: numerical integrations of the integral \Eq~(\ref{eq:tzlk}). Different colours correspond to different values of $\gamma$, indicated in the legend. Markers indicate results from numerical integrations of the equations of motion. The top and bottom panels correspond to $g_0=0.01$ rad and $g_0=1.5$ rad, respectively, whereas $e_0=0.01$ in both panels. The vertical coloured dotted lines show the critical value of $\gamma$ for the boundary between circulation and libration (cf. \Eq~\ref{eq:gammacrit}). }
\label{fig:t2set1}
\end{figure}

\begin{figure}
\center
\includegraphics[scale = 0.45, trim = 8mm 0mm 8mm 0mm]{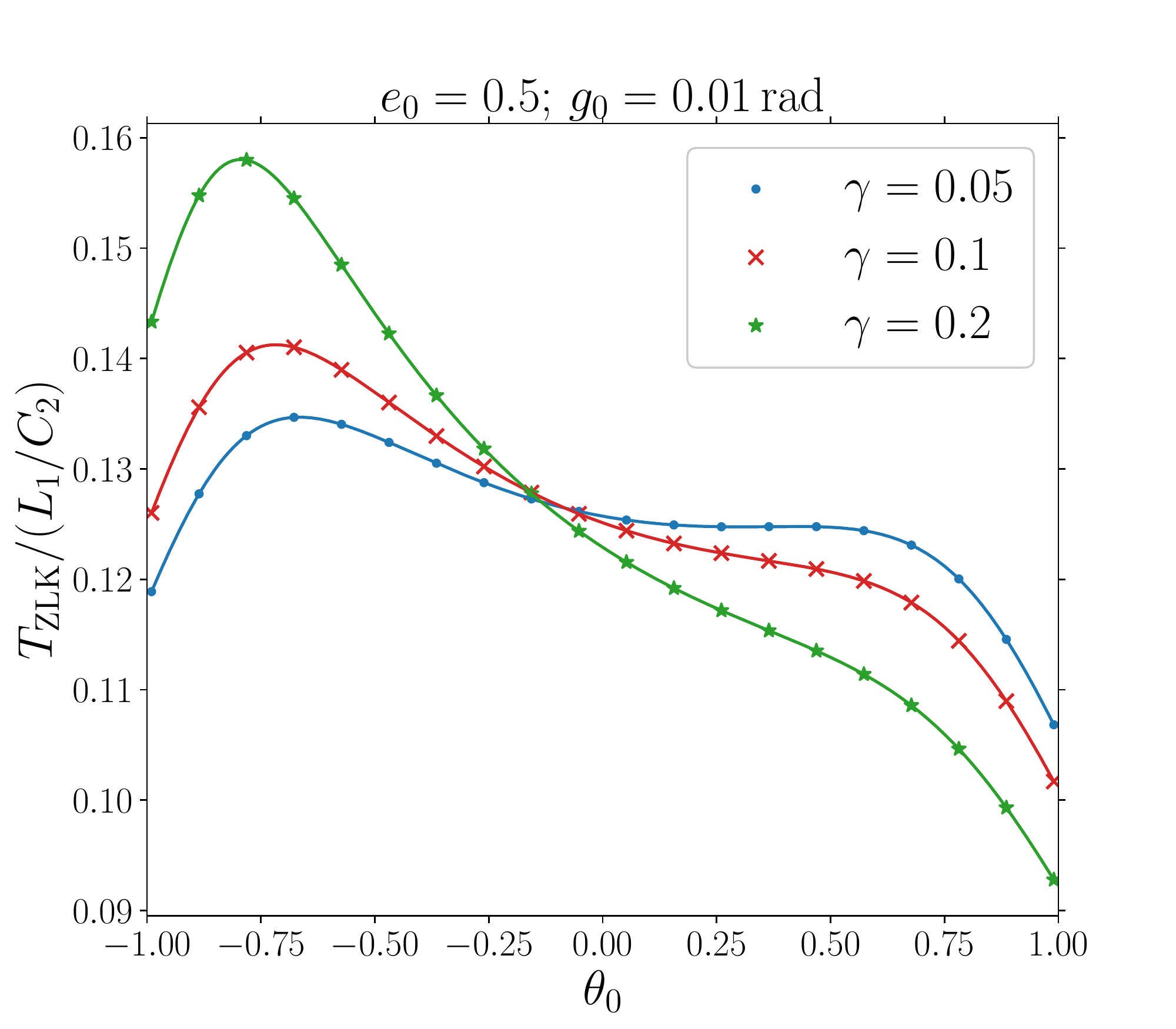}
\includegraphics[scale = 0.45, trim = 8mm 0mm 8mm 0mm]{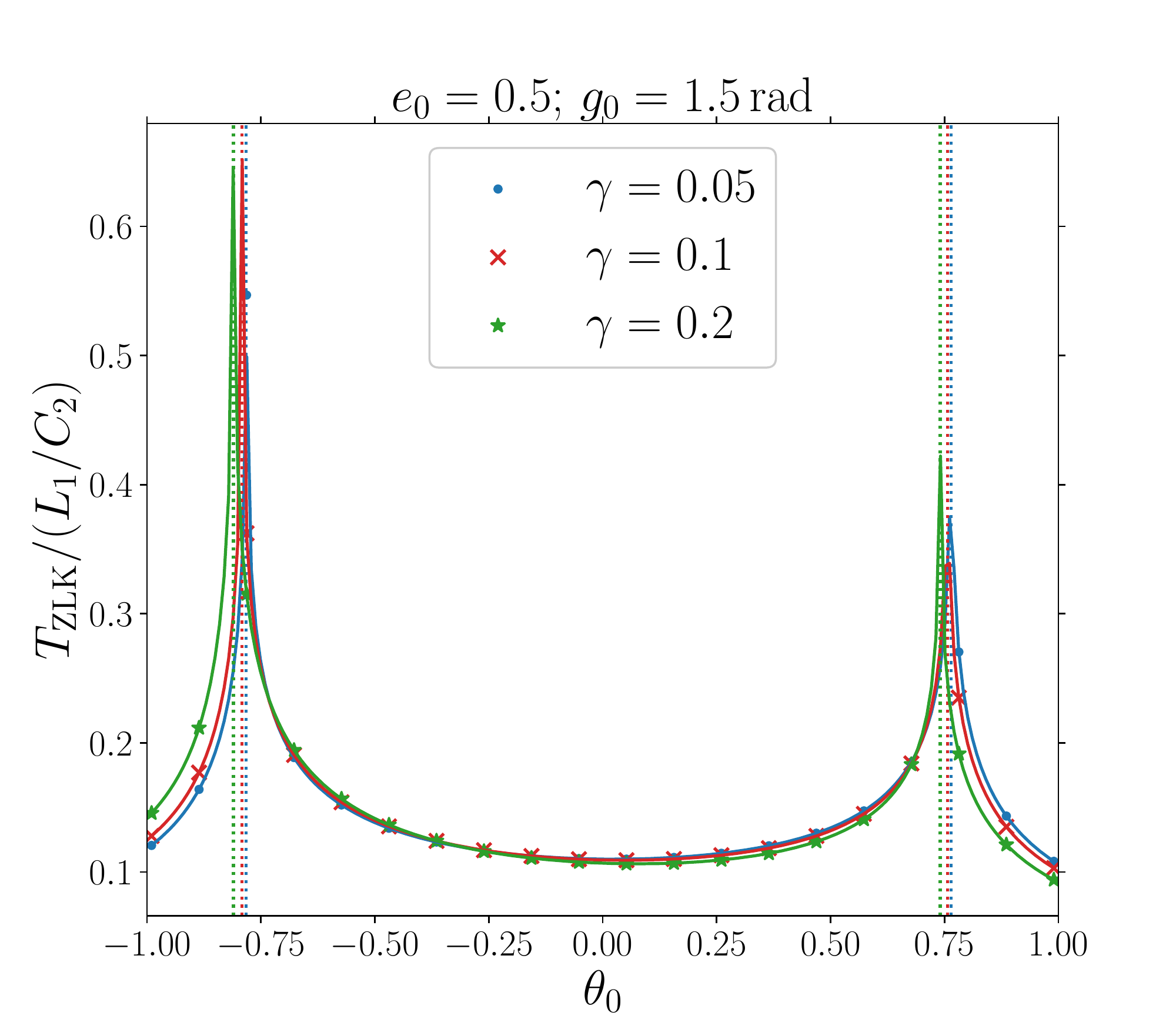}
\caption{Similar to \F~\ref{fig:t1set1}, now with a higher initial eccentricity, $e_0=0.5$. }
\label{fig:t1set2}
\end{figure}

\begin{figure}
\center
\includegraphics[scale = 0.45, trim = 8mm 0mm 8mm 0mm]{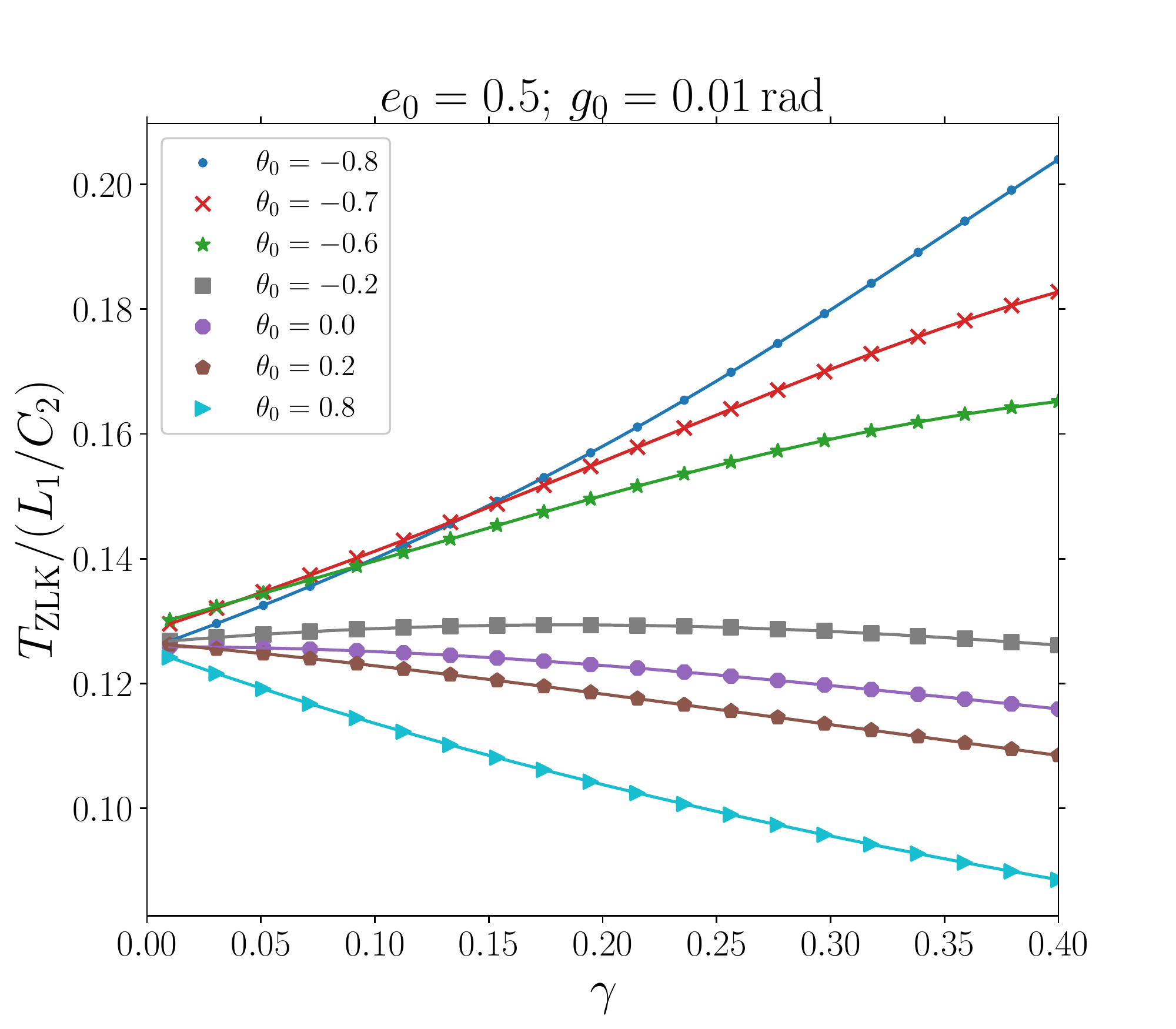}
\includegraphics[scale = 0.45, trim = 8mm 0mm 8mm 0mm]{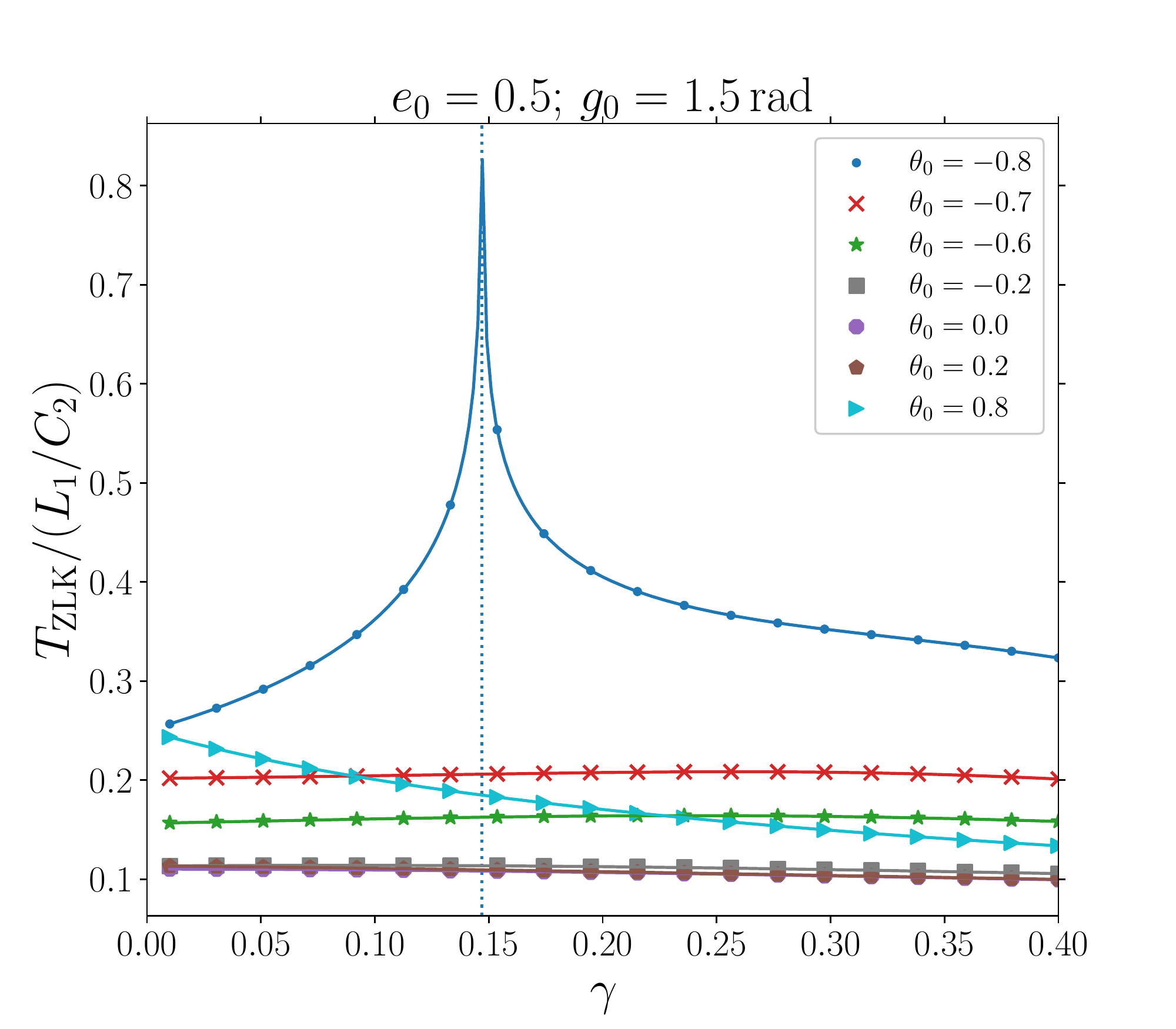}
\caption{Similar to \F~\ref{fig:t2set1}, now with a higher initial eccentricity, $e_0=0.5$. }
\label{fig:t2set2}
\end{figure}

\begin{figure}
\center
\includegraphics[scale = 0.45, trim = 8mm 0mm 8mm 0mm]{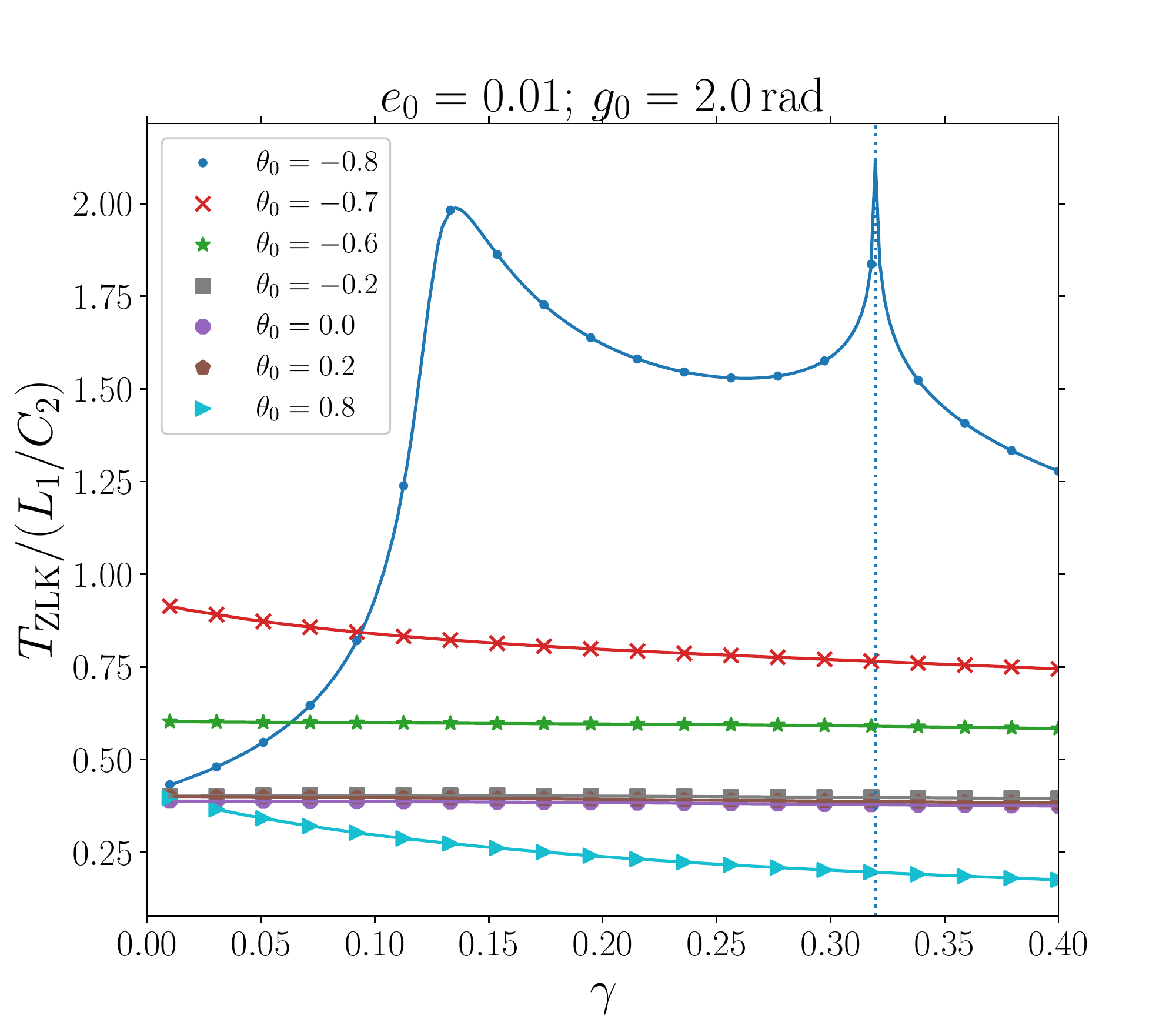}
\caption{Similar to \F~\ref{fig:t2set1} ($e_0=0.01$), but now with $g_0=2$ rad.}
\label{fig:t2set3}
\end{figure}

\subsection{Eccentricity oscillation timescales}
\label{sect:gamma:t}
For the same choice of parameters as in \S~\ref{sect:gamma:ecc}, we show in \Fs~\ref{fig:t1set1} through \ref{fig:t2set2} the timescales of eccentricity oscillations. The timescales are normalised to the quantity $L_1/C_2$ (cf. \Eq~\ref{eq:l1divc2}). Solid lines show numerical integrations of the integral \Eq~(\ref{eq:tzlk}), whereas markers indicate results from numerical integrations of the equations of motion (evidently, both methods require numerical calculation, but the latter is computationally more expensive). 

When considered as a function of $\theta_0$ (\Fs~\ref{fig:t1set1} and \ref{fig:t1set2}), $T_\ZLK$ shows distinct peaks near the `edges' $\theta_0= \pm1$. This is a result of the diverging timescales near the boundary between circulation and libration, and which has been pointed out by numerous authors before (e.g., \citealt{1999CeMDA..75..125K,2015MNRAS.452.3610A}). The locations of these boundaries in terms of $\theta_0$ can be calculated analytically from \Eq~(\ref{eq:cz}). They are shown with the vertical coloured dotted lines (where applicable), and match the locations where the timescales diverge according to the numerical calculations. In these plots of $e_\max$ as a function of $\theta_0$, changing $\gamma$ generally has no large impact. 

When considered as a function of $\gamma$ (\Fs~\ref{fig:t2set1} and \ref{fig:t2set2}), $T_\ZLK$ shows more complex behaviour which can be linked to the two effects described in \S~\ref{sect:gamma:ecc}. For small initial eccentricities (\F~\ref{fig:t2set1}), there is generally a weak dependence on $\gamma$. However, a maximum appears when $e_0=0.01$, $g_0=0.01$, and $\theta_0=-0.8$, near $\gamma=0.15$. This coincides with the eccentricity increase as a function of $\gamma$ in \F~\ref{fig:emax2set1} occurring for the corresponding parameters around $\gamma=0.15$. We associate the increase in $T_\ZLK$ in this case with the torque of the inner orbit on the outer orbit (cf. point ii of \ref{sect:gamma:ecc}), tending to increase the eccentricity timescale (see also \F~\ref{fig:time}, in which the eccentricity timescale in the case $\gamma=0.3$, $\simeq 0.45 \, L_1/C_2$, is slightly longer compared to the case $\gamma=0.05$, $\simeq 0.44 \, L_1/C_2$). 

In the bottom panel of \F~\ref{fig:t2set1} and focusing on the case $\theta_0=-0.8$, the local behaviour of $T_\ZLK$ as a function of $\gamma$ changes drastically compared to the top panel; instead of a smooth local maximum, a sharp peak appears near $\gamma=0.13$. The latter can be explained by the fact that $g_0$ affects the location of the boundary between circulation and libration; for $g_0=1.5$ (bottom panel of \F~\ref{fig:t2set1}), this boundary in terms of $\gamma$ (indicated with the vertical coloured dotted line, cf. \Eq~\ref{eq:gammacrit}), happens to be close to the local maximum associated with the inner orbit's torque. 

Furthermore, for different $g_0$, the circulation-libration boundary in $\gamma$ occurs at different locations, and it is possible that multiple maxima occur as a function of $\gamma$. We show an example of this in \F~\ref{fig:t2set3}, where $e_0=0.01$, and $g_0=2$ rad. The first local maximum around $\gamma=0.15$ is smooth and is associated with the inner orbital torque, whereas the second peak (around $\gamma=0.32$) is sharp and is associated with the circulation-libration boundary (vertical blue dotted line). 

For larger initial eccentricities (\F~\ref{fig:t2set2}), there is generally a more noticeable dependence on $\gamma$. For prograde orientations, $T_\ZLK$ decreases with increasing $\gamma$, which we associate with effect (i) of \S~\ref{sect:gamma:ecc}. For retrograde orientations, $T_\ZLK$ increases with increasing $\gamma$, which we associate with effect (ii) of \S~\ref{sect:gamma:ecc}.

\subsection{Orbital flips}
\label{sect:gamma:flip}

\begin{figure*}
\center
\includegraphics[scale = 0.55, trim = 8mm 0mm 8mm 0mm]{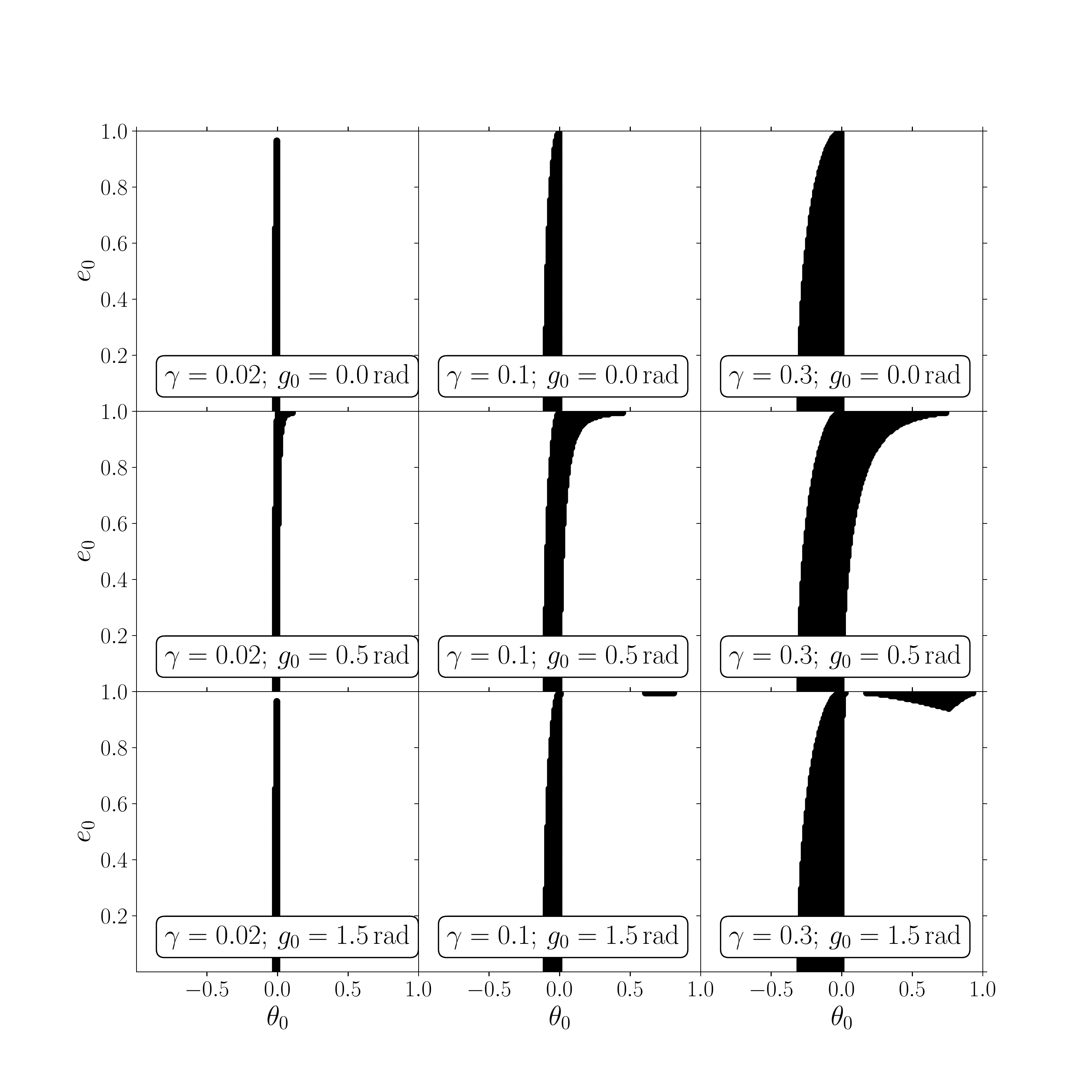}
\caption{Regions of the $(\theta_0,e_0)$ parameter space for which flips occur (see \Eq~\ref{eq:flipcrit}), indicated in black. Each panel corresponds to a different value of $g_0$ and $\gamma$, indicated therein. }
\label{fig:flip}
\end{figure*}

In addition to eccentricity excitation, it is of interest to consider excitation of the mutual inclination, $i_\rel$ (or, equivalently, $\theta = \cos i_\rel$). The analytic results from the previous sections can be used to determine whether orbital flips will occur for a given system, i.e., whether the mutual inclination angle will cross $i_{\rel} = 90^\circ$ at any point in the evolution. Such flips play an important role when, e.g., octupole-order effects are included (e.g., \citealt{2011ApJ...742...94L,2011PhRvL.107r1101K,2013ApJ...779..166T,2014ApJ...785..116L,2014ApJ...791...86L}), or more bodies are taken into account (e.g., \citealt{2013MNRAS.435..943P,2015MNRAS.449.4221H,2017MNRAS.470.1657H,2018MNRAS.474.3547G}). 

\begin{figure}
\center
\includegraphics[scale = 0.45, trim = 8mm 0mm 8mm 0mm]{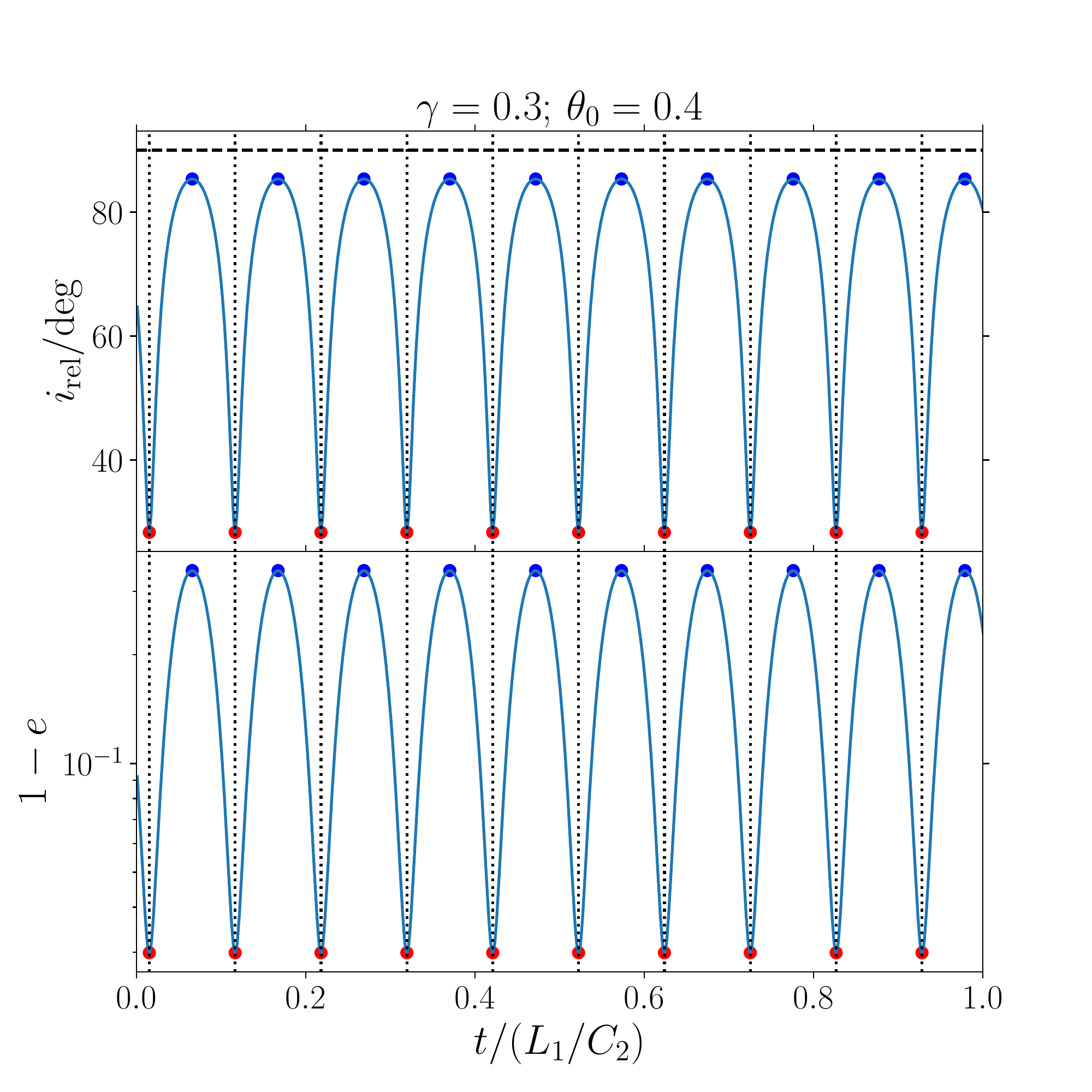}
\includegraphics[scale = 0.45, trim = 8mm 0mm 8mm 0mm]{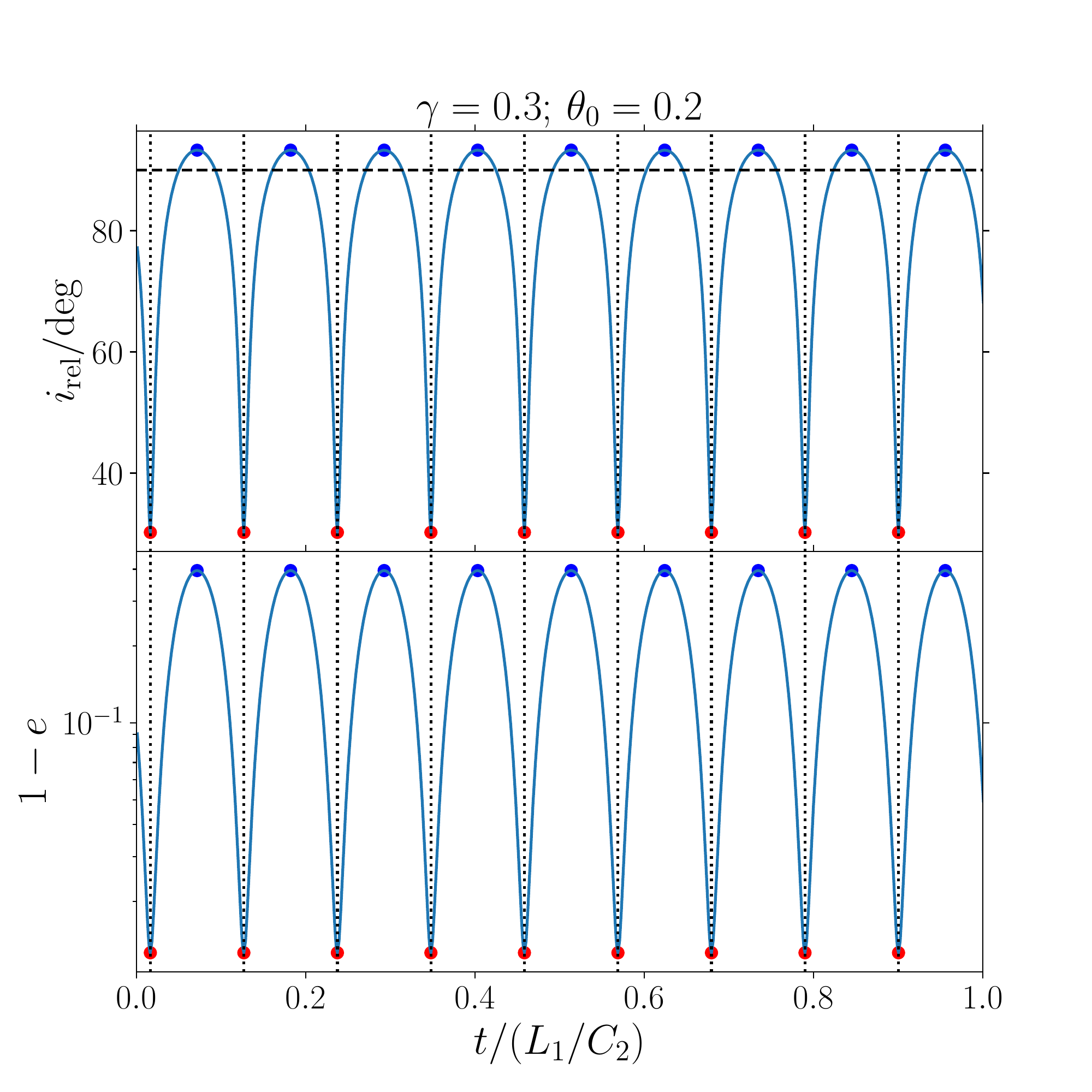}
\caption{Time evolution (from integration of the equations of motion) of the relative inclination (top panels) and the inner orbit eccentricity (bottom panels) for two systems with each $\gamma=0.3$, $g_0 = 0.5\,\mathrm{rad}$, and $e_0 = 0.9$. In the top and bottom set of panels, $\theta_0$ is taken to be 0.4 and 0.2, respectively. Blue and red dots indicate locations corresponding to eccentricity minima and maxima, respectively (determined via numerical root finding). Black vertical dotted lines also indicate eccentricity maxima. The black horizontal dashed lines in the top panels indicate $i_\rel=90^\circ$. Time is normalised to $L_1/C_2$ (cf. \Eq~\ref{eq:l1divc2}). }
\label{fig:time_flip}
\end{figure}

Specifically, from \Eq~(\ref{eq:theta}), it immediately follows that, if flips occur ($\theta=0$ at any time, with the understanding that $\theta_0 \neq 0$), the eccentricity at the moment of flip is given by
\begin{align}
\label{eq:xflip}
x_{\mathrm{flip}} = \frac{\TZ}{\gamma} = \sqrt{1-e_0^2} \left ( \sqrt{1-e_0^2} + \frac{\theta_0}{\gamma} \right ).
\end{align}
Note that $x_{\mathrm{flip}}$ could be negative or greater than unity, depending on $\TZ$ and $\gamma$. Since always $0<x\leq 1$, this implies that 
\begin{align}
\gamma > \frac{-\theta_0}{\sqrt{1-e_0^2}}; \qquad \gamma > \frac{\sqrt{1-e_0^2} \theta_0}{e_0^2}
\end{align}
is a necessary, but not guaranteed condition for flips to occur. Since, generally, $x_\min \leq x \leq x_\max$, the precise condition can be obtained by requiring that
\begin{align}
\label{eq:flipcrit}
x_\min \leq x_{\mathrm{flip}} \leq x_\max.
\end{align}
For given initial conditions ($\theta_0$, $e_0$, $g_0$, and $\gamma$), \Eq~(\ref{eq:flipcrit}) can be used in conjunction with \Eqs~(\ref{eq:xmingen}), (\ref{eq:xmaxgenlib}), and (\ref{eq:xmaxgencirc}) for $x_\min$ and $x_\max$ to determine whether or not orbital flips occur. We note that, in the test-particle limit $\gamma\rightarrow 0$, no flips can occur (in the latter case, $x_{\mathrm{flip}} \rightarrow \infty$ as shown by \Eq~\ref{eq:xflip}). 

In \F~\ref{fig:flip}, we show regions of the $(\theta_0,e_0)$ parameter space for which flips occur, indicated in black. Each panel corresponds to a different value of $g_0$ and $\gamma$. For small $\gamma$, the parameter space is small; flips only occur when $\theta_0$ is close to zero, and largely independent of $e_0$. For larger $\gamma$, the parameter space for flips increases. Flips occur for any $e_0$ if $\theta_0$ is slightly negative, and, depending on $g_0$, for high $e_0$ if $\theta_0$ is positive and sufficiently large. We note that similar behaviour, i.e., where orbits with initially high eccentricity and low inclination can flip, is manifested if octupole-order terms are taken into account \citep{2014ApJ...785..116L}.

As a demonstration, we show in \F~\ref{fig:time_flip} the time evolution for two systems with each $\gamma=0.3$, $g_0 = 0.5\,\mathrm{rad}$, and $e_0 = 0.9$. In the top and bottom set of panels, $\theta_0$ is taken to be 0.4 and 0.2, respectively. As can be inferred from the panel in the second row and third column in \F~\ref{fig:flip}, flips are not expected for these parameters if $\theta_0=0.4$, whereas they are if $\theta_0=0.2$. This is consistent with the behaviour shown in \F~\ref{fig:time_flip}. 

We note that the moment of flip is not necessarily associated with an eccentricity maximum (as is the case, for example, in \F~\ref{fig:time_flip}). It is straightforward to see from \Eq~(\ref{eq:xdot}) that, at $\theta=0$, 
\begin{align}
\dot{x} = \pm 4 \frac{C_2}{L_1} \sqrt{\frac{\TZ}{\gamma}} \sqrt{\left (20+c_0-18 \frac{\TZ}{\gamma} \right )\left(10-c_0 - 12 \frac{\TZ}{\gamma} \right )}.
\end{align}
This is not generally zero, implying that $\theta=0$ does not guarantee an eccentricity maximum.

\section{Discussion}
\label{sect:discussion}

\subsection{Maximum allowed value of $\gamma$}
\label{sect:discussion:gamma}
In the previous sections, we considered $\gamma$ to be a free parameter; in the quantitative investigation (\S~\ref{sect:gamma}), we restricted to $0<\gamma<0.4$. Generally, $\gamma$ cannot be increased arbitrarily. We showed that properties of the oscillations depend only on parameters of the system (masses, semimajor axes, and outer orbit eccentricity) through the quantity $\gamma$ (\Eq~\ref{eq:gammadef}). However, the allowed {\it range} of $\gamma$ is affected by the requirement of dynamical stability, and the latter does depend on such additional parameters. 

We can estimate the maximum allowed value of $\gamma$ for stability, $\gamma_\stab$, by using an analytic stability criterion. We adopt the criterion of \citet{2001MNRAS.321..398M}, i.e.,
\begin{align}
\label{eq:stab}
a_2 > \frac{a_1}{1-e_2} C_\stab \left [ \left (1+q_2 \right )\frac{1+e_2}{\sqrt{1-e_2}} \right  ]^{2/5}
\end{align}
for stability. Here, we neglect the relative inclination dependence, $C_\stab \equiv 2.8$, and $q_2 \equiv m_3/(m_1+m_2)$. The criterion of \citet{2001MNRAS.321..398M} applies well to systems with masses that are not too unequal. When large mass ratios are involved, other criteria are more appropriate (see, e.g., \citealt{2015ApJ...808..120P}). Here, for simplicity, we shall exclusively use \Eq~(\ref{eq:stab}).

Combining \Eqs~(\ref{eq:gammadef}) and (\ref{eq:stab}), it is straightforward to show that the largest value of $\gamma$ such that the system is still dynamically stable, is given by
\begin{align}
\label{eq:gammastab}
\gamma_\stab = \frac{1}{2 \sqrt{C_\stab}} \left ( \frac{\sqrt{1-e_2}}{1+e_2} \right )^{1/5} \frac{q_1}{(1+q_1)^2} \frac{\left(1+q_2 \right )^{3/10}}{q_2},
\end{align}
where $q_1 \equiv m_2/m_1$ ($0\leq q_1 \leq 1$, whereas $q_2>0$).

\begin{figure}
\center
\includegraphics[scale = 0.45, trim = 8mm 0mm 8mm 0mm]{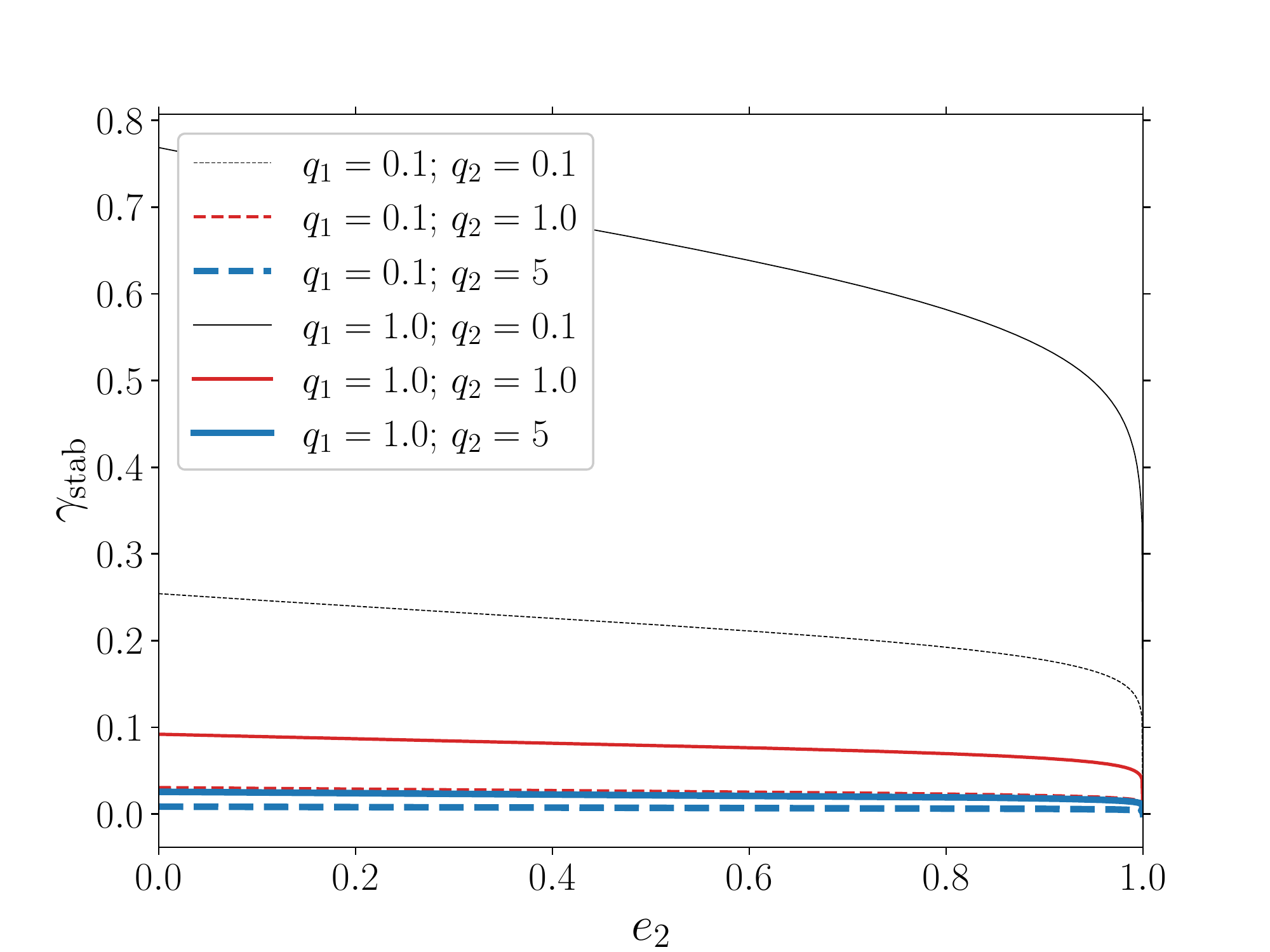}
\caption{Largest value of $\gamma$ allowed for dynamical stability (cf. \Eq~\ref{eq:gammastab}) according the criterion of \citet{2001MNRAS.321..398M}, plotted as a function of $e_2$ for different mass ratios $q_1 \equiv m_2/m_1$, and $q_2 \equiv m_3/(m_1+m_2)$. Refer to the legend for the meaning of the different colours and line styles.}
\label{fig:gammastab}
\end{figure}

In \F~\ref{fig:gammastab}, we plot \Eq~(\ref{eq:gammastab}) as a function of $e_2$ for different combinations of $q_1$ and $q_2$. Generally, the dependence on $e_2$ is weak, except for $e_2$ close to unity. There are strong dependences on the mass ratios; if $q_1=q_2=1$, the maximum allowed value of $\gamma$ with regard to stability is $\simeq 0.1$. This value decreases as $q_1$ is decreased, but increases significantly when $q_2$ is small. In particular, when $q_1=1$ and $q_2=0.1$, $\gamma_\stab \simeq 0.8$ for small $e_2$.

\subsection{Importance of the octupole-order terms}
\label{sect:discussion:oct}
A concern in addition to that of dynamical stability is the existence of higher-order expansion terms. It is well known that octupole-order terms can give rise to more complicated (even chaotic) behaviour compared to the quadrupole expansion order, and much higher eccentricities can be reached (e.g., \citealt{2011ApJ...742...94L,2011PhRvL.107r1101K,2013ApJ...779..166T,2014ApJ...785..116L,2014ApJ...791...86L}). The importance of the octupole terms can be quantified using the octupole parameter, which is composed of a ratio of leading factors in the octupole- and quadrupole-order expansion terms, i.e.,
\begin{align}
\label{eq:epsoct}
\epsoct = \frac{m_1-m_2}{m_1+m_2} \frac{a_1}{a_2} \frac{e_2}{1-e_2^2}.
\end{align}
Typically, $\epsoct \gtrsim 10^{-3}$ indicates that octupole-order terms are important. 

With \Eq~(\ref{eq:gammadef}), \Eq~(\ref{eq:epsoct}) can be written as
\begin{align}
\label{eq:epsoctgamma}
\epsoct = 4 \gamma^2 e_2 \frac{(1-q_1)(1+q_1)^3}{q_1^2} \frac{q_2^2}{1+q_2}.
\end{align}
Since $\gamma$ is also restricted by dynamical stability, the largest allowed value of $\epsoct$ is (cf. \Eq~\ref{eq:gammastab})
\begin{align}
\epsilon_{\mathrm{oct,\,\stab}} = \frac{e_2}{C_\stab} \left ( \frac{\sqrt{1-e_2}}{1+e_2} \right )^{2/5} \frac{1-q_1}{1+q_1} \left (1+q_2 \right )^{-2/5}.
\end{align}

\begin{figure}
\center
\includegraphics[scale = 0.45, trim = 8mm 0mm 8mm 0mm]{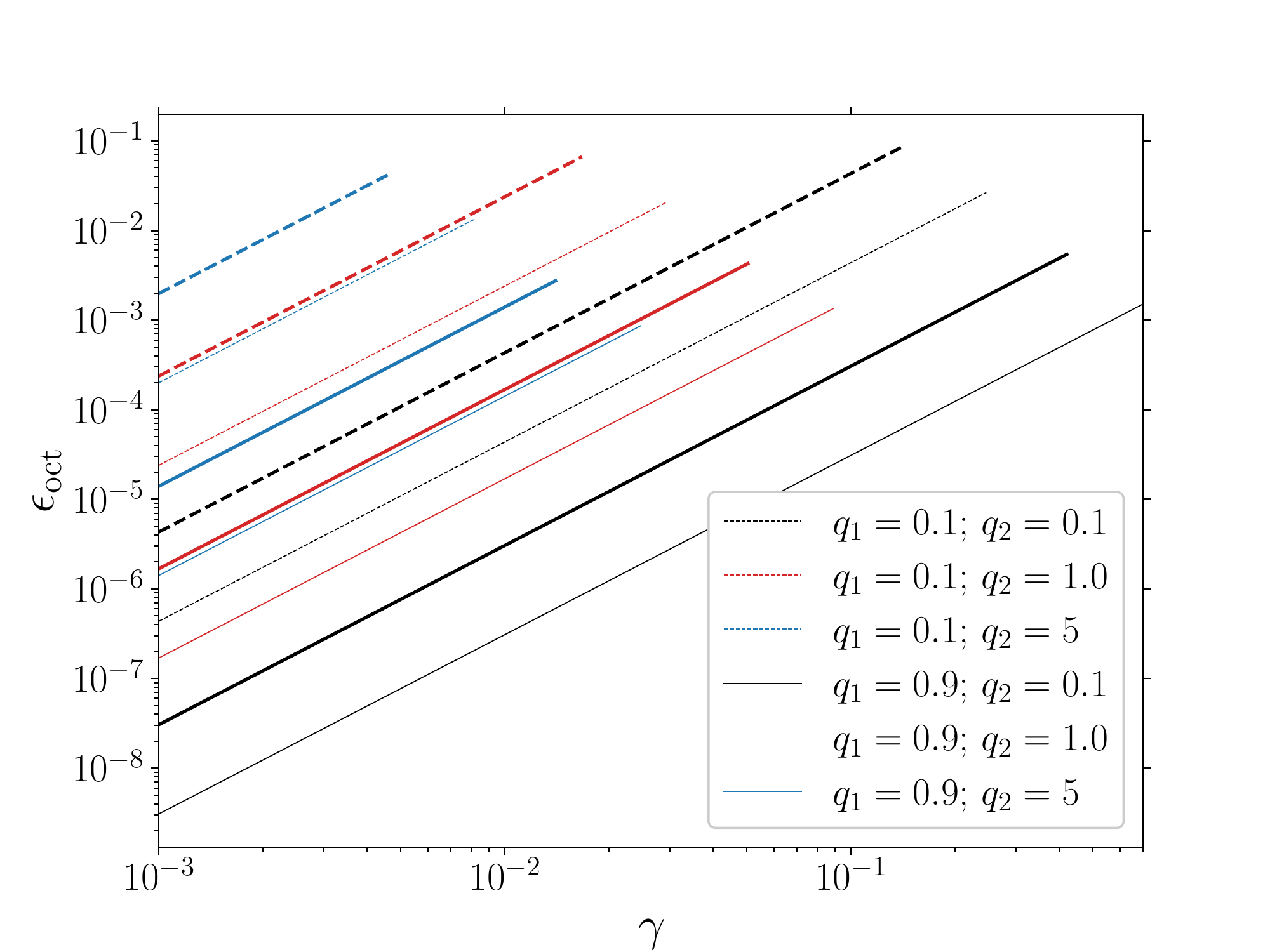}
\caption{Value of the octupole parameter, \Eq~(\ref{eq:epsoct}), as a function of $\gamma$ for several combinations of $q_1$, $q_2$, and $e_2$ (refer to the legend with regard to $q_1$ and $q_2$). Thin and thick lines correspond to $e_2 = 0.1$ and $e_2 = 0.99$, respectively. The range of $\gamma$ for each line is $10^{-3}<\gamma<\gamma_\stab$, where $\gamma_\stab$ is given by \Eq~(\ref{eq:gammastab}). }
\label{fig:gammaoct}
\end{figure}

In \F~\ref{fig:gammaoct}, we plot $\epsoct$ as a function of $\gamma$ in the range $10^{-3}<\gamma<\gamma_\stab$ for different combinations of $q_1$, $q_2$, and $e_2$ (note that $\epsoct$ in \Eq~\ref{eq:epsoctgamma} is directly proportional to $e_2$). In particular in systems with $q_1$ close to unity (solid lines in \F~\ref{fig:gammaoct}), $\epsoct$ is typically small and $<10^{-3}$, even for large $\gamma$. When $q_1$ is small, octupole-order terms become more important, particular when $e_2$ is large as well. Small $e_2$ increase the range of $\gamma$ for which $\epsoct$ is small.

\begin{figure}
\center
\includegraphics[scale = 0.45, trim = 8mm 0mm 8mm 0mm]{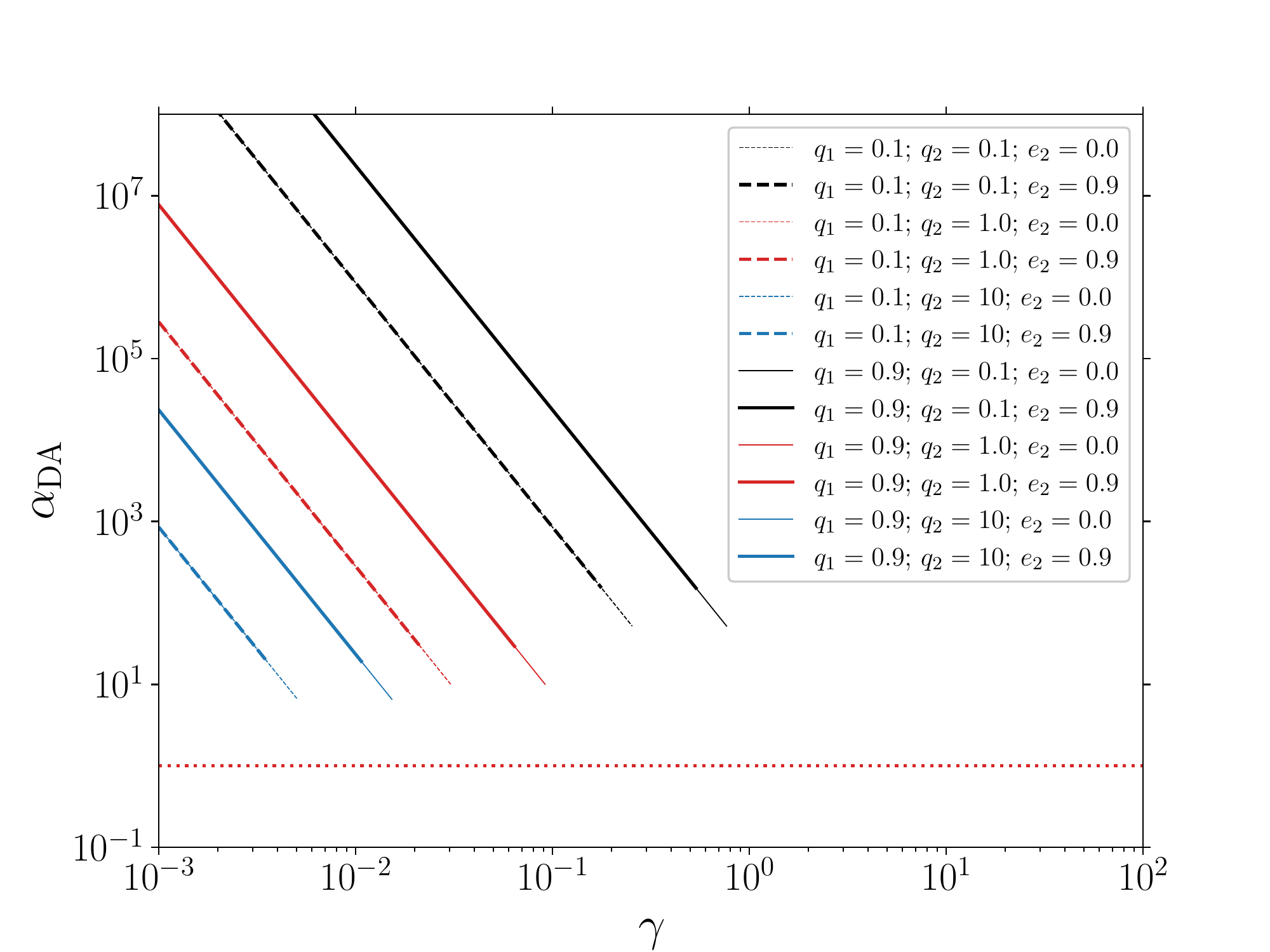}
\caption{Value of the quantity $\alpha_\DA$ defined in \Eq~(\ref{eq:alphada}) as a function of $\gamma$ for several combinations of $e_0$ (which affects $\gamma_\stab$), $q_1$, and $q_2$ (refer to the legend). The range of $\gamma$ for each line is $10^{-3}<\gamma<\gamma_\stab$, where $\gamma_\stab$ is given by \Eq~(\ref{eq:gammastab}). The red dotted horizontal line indicates $\alpha_\DA=1$, below which we expect the double averaging approximation to break down. }
\label{fig:gammada}
\end{figure}

\subsection{Double averaging approximation}
\label{sect:discussion:da}
Another concern is that the double averaging approximation, which underlies the Hamiltonian \Eq~(\ref{eq:H}), can break down when the timescale for angular momentum changes in the system becomes comparable to or even shorter than the orbital periods \citep{2012arXiv1211.4584K,2014MNRAS.438..573B,2014ApJ...781...45A,2014MNRAS.439.1079A,2016MNRAS.458.3060L,2018MNRAS.481.4907G,2018MNRAS.481.4602L,2019MNRAS.490.4756L,2020MNRAS.494.5492H}. We estimate the importance of the double averaging breakdown by comparing an approximation of the secular timescale (cf. \Eq~\ref{eq:l1divc2}) with the outer orbital period, i.e., we consider the quantity
\begin{align}
\label{eq:alphada}
\alpha_\DA \equiv \frac{P_2}{P_1} \frac{m_1+m_2+m_3}{m_3} \left (1-e^2 \right )^{3/2} = \frac{1}{8} \gamma^{-3} \frac{q_1^3(1+q_2)^2}{(1+q_1)^6 q_2^4}.
\end{align}
We expect the double averaging approximation to break down when $\alpha_\DA \lesssim 1$. 

In \F~\ref{fig:gammada}, we plot $\alpha_\DA$  as a function of $\gamma$ for $10^{-3}<\gamma<\gamma_\stab$, and several combinations of $e_0$ (which affects $\gamma_\stab$), $q_1$, and $q_2$. In most cases shown in the figure, $\alpha_\DA \gg 1$, and the double averaging is justified. However, in some cases, can be as small as $\sim 10$, indicating that non-secular effects may become important. Therefore, care should be taken when $\gamma$ is close to $\gamma_\stab$, not only from the point of view of dynamical stability, but also in terms of the validity of the averaging approximation.

\subsection{Distinguishing features}
\label{sect:discussion:dis}
Excitation of the orbital eccentricity can be due to a large range of mechanisms, not limited to the case considered here. For example, octupole-order effects can induce high eccentricities and orbital flips \citep{2011ApJ...742...94L,2011PhRvL.107r1101K,2013ApJ...779..166T,2014ApJ...785..116L,2014ApJ...791...86L}, and similar effects can occur in systems with more than three bodies (e.g., \citealt{2013MNRAS.435..943P,2015MNRAS.449.4221H,2017MNRAS.470.1657H,2018MNRAS.474.3547G}). In addition, eccentricity can be excited through scattering processes (e.g., \citealt{1996Sci...274..954R,2008ApJ...686..580C,2008ApJ...686..603J,2015ApJ...807...44P}). Based on the results presented above, we briefly comment that distinguishing features of non-test-particle interactions in hierarchical triples include:
\begin{itemize}[leftmargin=0.3cm]
\item A distribution of the maximum eccentricity (or, an associated property such as the number of mergers/strong interactions) which is asymmetric and not peaked at initial inclinations of $90^\circ$, but at slightly retrograde ones (see, e.g., \F~\ref{fig:emax1set1}). 
\item Eccentricity excitation in highly hierarchical triples (such that the quadrupole-order terms dominate) with highly retrograde initial inclinations. In the test-particle limit, no excitation would be expected for such systems, but this can change if $\gamma$ is sufficiently large (see, e.g., \F~\ref{fig:emax2set1}).
\item The occurrence of orbital flips, whereas these do not occur in the test-particle limit. However, flips are not necessarily associated with eccentricity maxima (see \S~\ref{sect:gamma:flip}). 
\end{itemize}

\subsection{Short-range forces}
\label{sect:discussion:srf}
Lastly, we briefly mention that, as is well known, apsidal motion induced by short-range forces due to, e.g., tidal bulges, rotation, or general relativity, can affect ZLK oscillations, usually reducing the maximum eccentricities (e.g., \citealt{2003ApJ...589..605W,2007ApJ...669.1298F,2013ApJ...773..187N,2015MNRAS.447..747L}). We did not consider these effects here since their inclusion would break the scale-invariance of the point mass Newtonian three-body problem, and the quantitative impact depends on the details of the short-range force. However, when applying our results to realistic systems, it should, of course, be taken into consideration that short-range forces could affect the maximum eccentricity.

\section{Conclusions}
\label{sect:conclusions}
We studied several properties of ZLK oscillations in hierarchical triple systems beyond the test particle approximation. Our results provide a deeper understanding of the behaviour of triples with bodies with comparable mass such as black hole and neutron star triple systems, and can be useful when interpreting results from numerical population synthesis studies. We showed that, when still limiting to the quadrupole expansion order, interesting behaviour can appear when the test particle approximation is relaxed. We quantified several properties as a function of $\gamma \equiv (1/2) \, L_1/G_2$, a ratio of inner-to-outer orbital angular momenta variables, where $\gamma=0$ in the test particle limit (cf. \Eq~\ref{eq:gammadef}). A \textsc{Python} script which implements our (semi)analytic expressions and numerical integration of the equations of motion is made freely available (see the link in \S~\ref{sect:gamma}).
 Our main conclusions are listed below.

\medskip \noindent 1. Based on both analytic and numerical methods, we considered in detail the dependence of the maximum eccentricities and eccentricity timescales as a function of $\gamma$ (\S~\ref{sect:gamma}). When considered as a function of $\theta_0\equiv \cos(i_{\rel,\,0})$, where $i_{\rel,\,0}$ is the initial relative inclination, we showed that eccentricity maximum occur around $\theta_0=-\gamma$ (the latter relation is exact when the initial eccentricity $e_0=0$). In other words, when the test particle approximation is relaxed, there exists symmetry breaking in the orientational dependence of secular eccentricity excitation. We interpret this as follows: when the initial orbital orientation is retrograde and $\gamma$ is sufficiently large, the inner orbit can torque the outer orbit and produce a relative inclination closer to $90^\circ$ (see, e.g., \F~\ref{fig:time}). This leads to more efficient eccentricity excitation for initially retrograde orientations. 

However, there is also a competing effect, which is that the inner orbit becomes less susceptible to torques from the outer orbit as the inner orbital angular momentum is increased in relative importance (i.e., increasing $\gamma$). For prograde orientations, this implies that the maximum eccentricity is reduced as $\gamma$ is increased. For (slightly) retrograde orbits, there exists a local maximum in $e_\max$ as a function of $\gamma$, which is given by $\gamma_\max=-\theta_0$ (valid in the limit $e_0=0$). 

\medskip \noindent 2. Typically, there is no strong dependence of the eccentricity timescale ($T_\ZLK$) on $\gamma$. Exceptions arise when the inner orbit is effective at torquing the outer orbit (effect ii discussed in \S~\ref{sect:gamma:ecc}), which can lead to smooth local maxima in $T_\ZLK$ as a function of $\gamma$. Another exception is when the solution lies on the boundary between circulation and libration. The latter boundary depends on $\gamma$, and can produce sharp peaks in $T_\ZLK$ as a function of $\gamma$. Therefore, it is possible that $T_\ZLK$ shows a complicated dependence as a function of $\gamma$, with multiple local maxima.

\medskip \noindent 3. In \S~\ref{sect:gen:ecc}, we presented closed-form analytic expressions for the minimum and maximum eccentricities as a function of $e_0$, $\theta_0$, $g_0$ (the initial inner orbit argument of periapsis), and $\gamma$. These expressions are exact within the limit of the quadrupole expansion order and within the double averaging approximation, i.e., they apply to any hierarchical triple as long as the system is dynamically stable, double averaging is justified, and effects due to octupole- and higher-order terms are negligible (we also assumed pure Newtonian point particles without short-range forces). We considered these conditions more quantitatively in \Ss~\ref{sect:discussion:gamma} through \ref{sect:discussion:da}.

\medskip \noindent 4. When $\gamma \neq 0$, phase-space trajectories in the $(\cos g,e)$ plane are shifted. The boundary between circulating and librating solutions is now dependent on $\gamma$, and given by the condition $\CZ=0$, where $\CZ$ is given by \Eq~(\ref{eq:cz}). The boundary can be formulated explicitly in terms of a critical value of $\gamma$ (as a function of $e_0$, $g_0$, and $\theta_0$), given by \Eq~(\ref{eq:gammacrit}). The fixed point condition, when the solution in the $(\cos g,e)$ plane reduces to a single point (i.e., no secular oscillations), is given by the $\gamma$-dependent \Eq~(\ref{eq:fpgen}). These properties reduce to the well known expressions in the test particle case ($\gamma \rightarrow 0$). 

\medskip \noindent 5. We also derived an analytic criterion for orbital flips to occur ($\theta$ crossing $\theta=0$ at any point in the evolution), and considered some of the associated parameter space (see \S~\ref{sect:gamma:flip}). 

\medskip \noindent 6. Given our approximations (stable hierarchical triple, double averaging, and quadrupole expansion order), when the test particle approximation is relaxed, all properties of the system can be quantified with the addition of only one parameter: $\gamma$. For example, the maximum eccentricity is unaffected when increasing the outer orbit eccentricity $e_2$, but decreasing the inner orbit semimajor axis $a_1$ in such a manner that $\gamma$ remains constant.

\section*{Acknowledgements}
I thank Patrick Neunteufel and Holly Preece for stimulating discussions and comments on the manuscript, and the anonymous referee for a helpful report. I also thank the Max Planck Society for support through a Max Planck Research Group.

\section*{Data availability}
The data underlying this article are available at \href{https://github.com/hamers/ZLK}{https://github.com/hamers/ZLK}.

\bibliographystyle{mnras}
\bibliography{literature}

\appendix


\label{lastpage}

\end{document}